\documentclass[a4paper, 12pt]{article}
\usepackage{amssymb}
\usepackage{amsmath,bm}
\usepackage{graphics}
\usepackage{epsfig}
\usepackage{subfigure}
\usepackage{cite}
\usepackage{multirow}
\usepackage[left=2cm, right=2cm]{geometry}
\newcommand{\tabincell}[2]{\begin{tabular}{@{}#1@{}}#2\end{tabular}}
\usepackage[font={footnotesize}]{caption}

\begin{document}
\bibliographystyle{unsrtnat}

\title{Analytic sensitivity analysis for models with correlated input variables}

\author{Yueying Zhu$^{1,2}$, Qiuping A Wang$^{1,3}$, Wei Li$^{2, 4}$, Xu Cai$^{4}$ \vspace{0.2cm}\\
\small $^1$ IMMM, UMR CNRS 6283, Le Mans Universit\'e, 72085 Le Mans, France\\
\small $^2$ Complexity Science Center \& Institute of Particle Physics,\\
\small Central China Normal University, 430079 Wuhan, China\\
\small $^3$ HEI, Yncrea, 59014 Lille, France\\
\small $^4$ Max-Planck Institute for Mathematics in the Sciences,\\
\small Inselst. 22, 04103 Leipzig, Germany\\
\small $^*$ Correspondence author: Yueying.Zhu.Etu@univ-lemans.fr
}

\date{}

\maketitle

\abstract{
An analytic formula is proposed to characterize the variance propagation from correlated input variables to the model response, by using multi-variate Taylor series. With the formula, partial variance contributions to the model response are then straightforwardly evaluated in the presence of input correlations. Additionally, an arbitrary variable is represented as the sum of independent and correlated parts. Universal expressions of the coefficients that specify the correlated and independent sections of a single variable are derived by employing linear correlation model. Based on the coefficients, it is nature to quantify the independent, correlated and coupling contributions to the total variance of model response. Numerical examples suggest the effectiveness and validation of our analytic framework for general models. A practical application of the analytic framework is also proposed to the sensitivity analysis of a deterministic HIV model.
}
\vspace{1cm}


\section{Introduction}
Uncertainty and sensitivity analysis is widely performed in various disciplines involving social science \cite{Mroz1987,Ligmann2014}, engineering science \cite{Becker2011}, economics \cite{Pannell1997}, chemistry \cite{Saltelli2005}, physics \cite{Carr1993,Morio2011,Timme2014,Posselt2016}, etc. It is quite useful for gaining insight into how input factors can be ranked according to their importance in establishing the uncertainty of model reponse. At present, many strategies have been built for the implementation of sensitivity analysis, including the traditional approach of changing one factor at a time \cite{Guo2013,Zhu2013}, local method \cite{Cacuci2005,Frankel2010,Farina2013,Pedro2016}, regression analysis \cite{Iooss2015}, variance-based method \cite{zhu2017}, etc. Among the various available strategies, variance-based sensitivity analysis has been assessed as versatile and effective for uncertainty and sensitivity analysis of complex models. The consideration of variance-based importance measures can be traced back to over twenty years ago when Sobol' characterized the first-order sensitivity measures on the basis of deposing the variance in model response into different partial contributions attributable to individual input variables and to their combinations (called variance decomposition) \cite{Sobol1993}. Then extensive relevant investigations are carried out around this Sobol's work, boiling down to the improvements in analysis strategies and to their applications to the sensitivity and reliability analysis of complex systems \cite{Helton2006,Kucherenko2015}. However, these frameworks are often proposed when the input variables are assumed to be statistically independent of each other.

Recently, the interest in extending sensitivity analysis strategies from uncorrelated case to the correlated one is increasing due to the existence of correlated input factors in practical applications. Previous investigations about sensitivity analysis with correlated input variables only provided overall sensitivity indices with respect to individual factors. However, the correlated and independent variance contributions were absent \cite{Saltelli2001}. In practical applications, the distinction between independent and correlated variance contributions is quite important. It allows one to decide whether the correlations among input factors should be considered or not.

Both the correlated and independent variance contributions were firstly considered by C. Xu {\it et al} \cite{Xu2008}. They proposed a regression-based strategy to decompose partial variance contributions into independent and correlated parts assuming linear relationship connecting the model response and input variables. To overcome the limitation of their method, many frameworks on sensitivity analysis are recently developed in the presence of correlated inputs, contraposing the investigation of more effective and universal technics for sensitivity analysis in general correlated situations\cite{Most2012,Hao2013,Li2017}. Still, a theoretical framework for the determination of partial variance contributions and of relative effects contributed by the independence and correlations of input variables is limited, especially when a single input is correlated with many others simultaneously.

In this work, an analytic formula is firstly derived to characterize the variance propagation from correlated inputs to the model response. In its implementation, the partial derivative of model response evaluated at central point of input vector and the covariance among different inputs are involved. They both provide fruitful information for evaluating partial variance contributions produced by individual variables alone and also by their interactions to the model response. With linear correlation model, we represent a single variable as the sum of independent and correlated sections. Universal expressions are then proposed for the coefficients which specify the independent and correlated sections of a single input by given pairwise correlations. The coefficients serve to identify the sensitivity of model response with respect to both the independent and correlated sections of individual input factors. Furthermore, except for the independent and correlated variance contributions, the partial variance produced by the coupling effect between independence and correlations of input variables can be also identified from the total variance of model response.

The rest of the paper is organized as follows. In section 2, the analytic formula for variance propagation is generalized to a general case with correlated input variables. The independent, correlated and coupling variance contributions are also interpreted in this section. Sensitivity measures are defined in section 3, in the presence of input correlations. The generation process of correlated variables is also analyzed here, by using linear correlation model. In section 4, four fabricated numerical models illustrate the effectiveness and applicability of our analytic framework, accompanied by a practical application to the sensitivity analysis of a deterministic HIV model. Section 5 gives concluding remarks.

\section{variance propagation}
Any operation that we perform on a model response dependent upon variables of uncertainty requires us to identify the response uncertainty based on the uncertainty in input variables. The propagation of variance, characterizing the effect of input uncertainty on the uncertainty of model response, constitutes the essential ingredient of uncertainty and sensitivity analysis of complex models.

\subsection{Independent case}
Consider a general mathematical model of the form $y=f(\bm{x})$ with $\bm{x}=(x_1, x_2, \cdots, x_n)^{\rm T}$ labeling the input vector of $n$-dimensional variables of uncertainty. The Taylor series of model response $y$ at the center point of input vector is represented as
\begin{equation}
\label{taylorgeneral}
\begin{split}
y=&f(\{\mu\})+\sum_{t=1}^{n}\sum_{i=1}^{\infty}\frac{1}{i!}(\frac{\partial^i f}{\partial x_t^{i}})(\{\mu\})\cdot(x_t-\mu_t)^{i}\\
&+\sum_{\substack{t,s=1;
t<s}}^{n}\sum_{i_t,i_s=1}^{\infty}\frac{1}{i_t!\cdot i_s!}(\frac{\partial^{i_t+i_s}f}{\partial x_t^{i_t}\partial x_s^{i_s}})(\{\mu\})\cdot (x_t-\mu_t)^{i_t}(x_s-\mu_s)^{i_s}+\cdots\\
&+\sum_{i_1\cdots i_{n}}^{\infty}\frac{1}{i_1!\cdots i_{n}!}(\frac{\partial^{i_1+\cdots+i_{n}}f}{\partial x_1^{i_1}\cdots\partial x_{n}^{i_{n}}})(\{\mu\})\cdot (x_1-\mu_1)^{i_1}\cdots (x_{n}-\mu_{n})^{i_{n}},
\end{split}
\end{equation}
where $\{\mu\}$ indicates the mathematical expectation set of input variables: $\{\mu_1, \mu_2, \cdots, \mu_{n}\}$. By definition, the universal expression of the variance of model response is derived, see Ref.\cite{zhu2017}, as
\begin{equation}
\label{vuncorrelated}
\begin{split}
V(y)=&\sum_{\substack{i_1\cdots i_{n}=0,\\
j_1\cdots j_{n}=0}}^{\infty}\frac{1}{A(i_1,\cdots, i_{n}, j_1,\cdots, j_{n})}\left(\frac{\partial^{i_1+\cdots+i_{n}}f}{\partial x_1^{i_1}\cdots \partial x_{n}^{i_{n}}}\cdot \frac{\partial^{j_1+\cdots+j_{n}}f}{\partial x_1^{j_1}\cdots \partial x_{n}^{j_{n}}}\right)(\{\mu\})\\
&\times F_{(x_1)^{i_1j_1}\cdots (x_{n})^{i_{n}j_{n}}},
\end{split}
\end{equation}
where input variables are assumed to be statistically independent of each other, and
\begin{align}
&A(\cdots)=i_1!\cdots i_{n}!\cdot j_1!\cdots j_{n}!\label{factorial},\\
&F_{(x_1)^{i_1j_1}\cdots (x_{n})^{i_{n}j_{n}}}=M_{i_1+j_1}(x_1)\cdots M_{i_{n}+j_{n}}(x_{n})-M_{i_1}(x_1)M_{j_1}(x_1)\cdots M_{i_{n}}(x_{n})M_{j_{n}}(x_{n}).\label{funcorrelated}
\end{align}
$M_i(x_j)$ indicates the $i^{th}$ central moment of variable $x_j$. Its general expressions are presented in Appendix \ref{appendixcentral} for both uniform and normal distributions. Equation (\ref{vuncorrelated}) contains fractional contributions of different dimensionality attributable to individual input variables and to their combinations. It allows one to analytically obtain the summands of increasing dimensionality that appeared in the variance decomposition of model response,
\begin{equation}
\label{vduncorrelated}
V(y)=\sum_{i=1}^{n}V_i+\sum_{\substack{i,j=1,
i<j}}^{n}V_{ij}+\sum_{\substack{i,j,k=1,
i<j<k}}^{n}V_{ijk}+\cdots+V_{12\cdots n},
\end{equation}
in which, $V_i$ is the first-order (or main) variance contribution produced by $x_i$ alone, $V_{ij}$ the contribution of the interaction associated with $x_i$ and $x_j$, and so on up to the last term the contribution of the interaction involving all input variables \cite{Sobol1993}.

\subsection{Correlated case}

Regarding the presence of correlated input variables, the analytic expression of variance propagation, see Eq. (\ref{vuncorrelated}), should be extended from independent case to the correlated one. Recalling the multi-variate Taylor series at the central point of input vector, presented in Eq. (\ref{taylorgeneral}) , the mathematical expectation of the model response can be represented in the presence of input correlations as
\begin{equation}
\label{ey}
\begin{split}
E(y)=&f(\{\mu\})+\sum_{t=1}^n\sum_{i_t=1}^{\infty}\frac{M_{i_t}(x_t)}{A(i_t)}(\frac{\partial^{i_t}f}{\partial x_t^{i_t}})(\{\mu\})+\sum_{\substack{t,s=1\\t<s}}^n\sum_{i_t,i_s=1}^{\infty}\frac{{\rm cov}(x_t^{i_t}, x_s^{i_s})}{A(i_t, i_s)}(\frac{\partial^{i_t+i_s}f}{\partial x_t^{i_t}\partial x_s^{i_s}})(\{\mu\})\\
&+\cdots+\sum_{i_1,\cdots, i_n=1}^{\infty}\frac{{\rm cov}(x_1^{i_1},\cdots, x_n^{i_n})}{A(i_1, \cdots, i_n)}(\frac{\partial^{i_1+\cdots+i_{n}}f}{\partial x_1^{i_1}\cdots \partial x_{n}^{i_{n}}})(\{\mu\})\\
=&\sum_{i_1,\cdots, i_n=0}^{\infty}\frac{{\rm cov}(x_1^{i_1},\cdots, x_n^{i_n})}{A(i_1, \cdots, i_n)}(\frac{\partial^{i_1+\cdots+i_{n}}f}{\partial x_1^{i_1}\cdots \partial x_{n}^{i_{n}}})(\{\mu\}),
\end{split}
\end{equation}
where ${\rm cov}(\cdot)$ is the covariance introduced by the correlations among input variables. It is defined as
\begin{equation}
\label{highecovariance}
{\rm cov}(x_1^{i_1}, \cdots, x_n^{i_n})=\int(x_1-\mu_1)^{i_1}\cdots (x_n-\mu_n)^{i_n}P(\bm{x}) {\rm d} \bm{x},
\end{equation}
where $P(\bm{x})$ is the joint probability density function of $\bm x$. In the absence of input correlations, $P(\bm{x})$ can be simplified as
\begin{equation}
P(\bm{x})=\prod_{i=1}^nP(x_i),
\end{equation}
in which $P(x_i)$ denotes the probability density function of input $x_i$. In the existence of correlations among input variables, the variance of model response is derived by the use of Eqs. (\ref{taylorgeneral}) and (\ref{ey}) as
\begin{multline}
\label{vcorrelated}
V(y)=\sum_{\substack{i_1,\cdots, i_{n}=0\\
j_1,\cdots, j_{n}=0}}^{\infty}\frac{1}{A(i_1,\cdots, i_{n}, j_1,\cdots, j_{n})}\left(\frac{\partial^{i_1+\cdots+i_{n}}f}{\partial x_1^{i_1}\cdots \partial x_{n}^{i_{n}}}\cdot \frac{\partial^{j_1+\cdots+j_{n}}f}{\partial x_1^{j_1}\cdots \partial x_{n}^{j_{n}}}\right)(\{\mu\})\\
\times \left[{\rm cov}(x_1^{i_1+j_1}, \cdots, x_n^{i_n+j_n})-{\rm cov}(x_1^{i_1},\cdots, x_n^{i_n})\cdot{\rm cov}(x_1^{j_1},\cdots, x_n^{j_n})\right].
\end{multline}
with $A(\cdots)$ defined in Eq. (\ref{factorial}). Apparently, the above generalized expression will be degenerated into Eq. (\ref{vuncorrelated}) if input variables are assumed to be statistically independent of each other.

The concept of complete variance decomposition, presented in Eq. (\ref{vduncorrelated}), is proposed by assuming input independence. Its form is also valid for the correlated case. In the presence of correlated input variables, however, the partial variance contributions with dimensionality larger than 1 is contributed not only by the coupling items presented in the functional form of the model under discussion (for independent case), but also by the input correlations. Regarding the situation of correlated inputs, the impact of a single variable can be represented as the sum of the contribution made by its correlations with the remaining variables and that made by its independence. Based on the description, each fractional variance contribution included in the original variance decomposition can be divided into three sections: independent variance contribution (labeled by superscript U), correlated variance contribution (labeled by superscript C), and coupling variance contribution (labeled by superscript UC). Mathematically, the output variance is decomposed in the presence of input correlations as
\begin{equation}
\label{vdcorrelated}
V(y)=\sum_{i=1}^{n}(V_i^{\rm U}+V_i^{\rm C}+V_i^{\rm UC})+\sum_{\substack{i,j=1;
i<j}}^{n}(V_{ij}^{\rm U_{p}}+V_{ij}^{\rm C_{p}}+V_{ij}^{\rm UC_{p}})+\cdots+(V_{12\cdots n}^{\rm U_q}+V_{12\cdots n}^{\rm C_q}+V_{12\cdots n}^{\rm UC_q}),
\end{equation}
where $p\in \{i, j\}$, $q\in \{1, 2, \cdots, n\}$, and
\begin{align}
&V_i=V_i^{\rm U}+V_i^{\rm C}+V_i^{\rm UC},\\
&V_{ij}=V_{ij}^{{\rm U}_{p}}+V_{ij}^{{\rm C}_{p}}+V_{ij}^{{\rm UC}_{p}},\\
&\vdots\nonumber\\
&V_{12\cdots n}=V_{12\cdots n}^{{\rm U}_q}+V_{12\cdots n}^{{\rm C}_q}+V_{12\cdots n}^{{\rm UC}_q}.
\end{align}
$V_i^{\rm U}$ ($V_i^{\rm C}$) is the variance contribution produced by the independent (correlated) section of $x_i$ alone, $V_{ij}^{{\rm U}_i}$ ($V_{ij}^{{\rm C}_i}$) the contribution of the interaction between $x_j$ and the independent (correlated) section of $x_i$, and so on up to $V_{12\cdots n}^{{\rm U}_q}$ ($V_{12\cdots n}^{{\rm C}_q}$) the contribution of the interaction associated with the independent (correlated) section of $x_q$ and the rest variables. Coupling variance contributions are produced by the coupling effects between independent and correlated sections of individual input variables.

\section{Estimation of sensitivity indices}
Working within a probabilistic framework, variance-based sensitivity measures are defined on the bases of partial contributions presented in the variance decomposition of model response. In the determination of each partial variance contribution, high-order covariance embodied in the analytic formula Eq. (\ref{vcorrelated}) should be concerned for nonlinear models. Consequently, it is necessary to specify the correlated and independent parts of single input variables for the confirmation of fractions contained in Eq. (\ref{vdcorrelated}), whereby the importance of independence, correlation, and the coupling between them can be quantified for individual input variables in establishing the uncertainty of model response.

\subsection{Generation of correlated variables}

In probabilistic models the dependency between two variables is often represented by the Pearson correlation coefficient which indicates pairwise linear correlations:
\begin{equation}
\label{linearcorrelation}
\rho(x_i, x_j)=\frac{E[(x_i-\mu_i)(x_j-\mu_j)]}{\sigma_i\sigma_j},
\end{equation}
with mean values $\mu_i$, $\mu_j$ and standard deviations $\sigma_i$, $\sigma_j$. $E[\star]$ is the expectation operation by returning the average value of $\star$. For the sake of simplicity in writing, $\rho(x_i, x_j)$ is simplified as $\rho_{ij}$ in the following discussion, that indicates the Pearson correlation coefficient between input variables $x_i$ and $x_j$.

In the presence of correlations, an arbitrary variable $x_i$ can be represented as the sum of a correlated section and an independent section:
\begin{equation}
x_i=x_i^{\rm C}+x_i^{\rm U}.
\end{equation}
The correlated section $x_i^{\rm C}$ indicates the correlations of $x_i$ with the remaining input variables. By the use of linear correlation model (Eq. (\ref{linearcorrelation})), $x_i^{\rm C}$ can be generated by a linear combination of the rest variables:
\begin{equation}
x_i^{\rm C}=\sum_{j=1,j\neq i}^na_{ij}x_j.
\end{equation}
The independent section $x_i^{\rm U}$ denotes the independence of $x_i$. It is often specified by a newly introduced random variable $r_i$:
\begin{equation}
x_i^{\rm U}=c_ir_i,
\end{equation}
in keeping the mean value $\mu(r_i)=(\mu_i-\sum_{j=1, j\neq i}^na_{ij}\mu_j)/c_i$ and standard deviation $\sigma(r_i)=\sigma_i$. Coefficients $\{a_{ij}, c_i; i, j=1, 2, \cdots, n\}$ specify the correlated and independent sections of $x_i$. They are determined by given pairwise correlations through equations
\begin{multline}
a_{ij}=\frac{\sigma_i}{\sigma_j}\left[1-\sum_{\substack{k<l,\\
k,l\neq i}}\rho_{kl}^2(1-\sum_{\substack{h<q,\\h>k;h,q\neq i,l}}\rho_{hq}^2)+2\sum_{\substack{k<l<h,\\
k,l,h\neq i}}\rho_{kl}\rho_{kh}(\rho_{lh}-\sum_{q\neq i,l,h;q>k}\rho_{lq}\rho_{hq})\right]^{-1}\\
\times \left[\rho_{ij}(1-\sum_{\substack{k<l,\\
k,l\neq i,j}}\rho_{kl}^2+2\sum_{\substack{k<l<h,\\
k,l,h\neq i,j}}\rho_{kl}\rho_{kh}\rho_{lh})-\sum_{\substack{
k\neq i,j}}\rho_{ik}\rho_{jk}(1-\sum_{\substack{h<q,\\
h,q\neq i,j,k}}\rho_{hq}^2)\right.\\
\left.+\sum_{\substack{k<l,\\
k,l\neq i,j}}(\rho_{ik}\rho_{jl}+\rho_{il}\rho_{jk})(\rho_{kl}-\sum_{h\neq i,j,k,l}\rho_{kh}\rho_{lh})\right]
,\label{coefficientcorrelated}
\end{multline}
and
\begin{multline}
c_i=\left[1-\sum_{\substack{k<l,\\
k,l\neq i}}\rho_{kl}^2(1-\sum_{\substack{h<q,\\h>k;
h,q\neq i,l}}\rho_{hq}^2)+2\sum_{\substack{k<l<h,\\
k,l,h\neq i}}\rho_{kl}\rho_{kh}(\rho_{lh}-\sum_{q\neq i,l,h;q>k}\rho_{lq}\rho_{hq})\right]^{-1/2}\\
\times\left[1-\sum_{k<l}\rho_{kl}^2(1-\sum_{\substack{h<q,\\ h>k;h,q\neq l}}\rho_{hq}^2)+2\sum_{k<l<h}\rho_{kl}\rho_{kh}(\rho_{lh}-
\sum_{q\neq l,h;q>k}\rho_{lq}\rho_{hq})\right]^{\frac{1}{2}},\label{coefficientuncorrelated}
\end{multline}
where high-order ($\geq 5$) terms are neglected. The expressions above, derived according to the analysis of simple cases as shown in Appendix \ref{appendixlinearcorrealtion}, constitute essential ingredients for the quantification of sensitivity measures associated with correlated section, independent section, and their coupling, for each single input variable.

\subsection{Sensitivity indices}
With help of the analytic formula Eq. (\ref{vcorrelated}) that explains the variance propagation in the presence of input correlations, the partial variance contributions of different dimensionality can be calculated by
\begin{align}
V_i=&\sum_{k, l=0}^{\infty}\frac{1}{k!\cdot l!}(\frac{\partial ^kf}{x_i^k}\cdot \frac{\partial ^lf}{\partial x_i^l})(\{\mu\})\cdot\left [M_{k+l}(x_i)-M_k(x_i)M_l(x_i)\right ], \label{firstorder}\\
V_{ij}=&\sum_{k,l,p,q=0}^{\infty}\frac{1}{k!\cdot l!\cdot p!\cdot q!}(\frac{\partial ^{k+p}f}{x_i^kx_j^p}\cdot \frac{\partial ^{l+q}f}{\partial x_i^lx_j^q})(\{\mu\})\cdot\left[{\rm cov}(x_i^{k+l}, x_j^{p+q})-{\rm cov}(x_i^k, x_j^p){\rm cov}(x_i^l, x_j^q)\right]\nonumber\\
&-V_i-V_j, \label{secondorder}\\
\vdots\nonumber
\end{align}
The high-order covariance between two variables, contained in Eq. (\ref{secondorder}), is defined (simplified from Eq. \ref{highecovariance}) as
\begin{equation}
\label{high-order-covariance}
{\rm cov}(x_i^k, x_j^l)=\int (x_i-\mu_i)^k (x_j-\mu_j)^l P(x_i, x_j) {\rm d}x_i{\rm d}x_j,
\end{equation}
which acts as a function of the Pearson correlation coefficient between $x_i$ and $x_j$. ${\rm cov}(x_i^k, x_j^l)$ can be derived analytically by formulating one variable on the basis of another:
\begin{eqnarray}
x_i&=\rho_{ij}\frac{\sigma_i}{\sigma_j}x_j+\sqrt{1-\rho_{ij}^2}r_i,\\
x_j&=\rho_{ij}\frac{\sigma_j}{\sigma_i}x_i+\sqrt{1-\rho_{ij}^2}r_j,
\end{eqnarray}
with $r_i$ ($r_j$) independent of $x_j$ ($x_i$) and having the same variance as $x_i$ ($x_j$). If $k\neq l$, the above two formulating strategies are equivalent in determining ${\rm cov}(x_i^k, x_j^l)$ only when both $x_i$ and $x_j$ are normally distributed. Furthermore, if one focuses on the correlated, independent and coupling effects contained in the hight-order covariance Eq. (\ref{high-order-covariance}), $x_i$ ($x_j$) should be formulated on the bases of all those variables which might be correlated with $x_i$ ($x_j$).

The total contributions to the variance of model response, associated with independent, correlated, and coupling effects are represented, for an arbitrary variable $x_i$, as
\begin{align}
&V_i^{\rm TU}=V_i^{\rm U}+\sum_{j\neq i}^n V_{ij}^{{\rm U}_i}+\cdots+V_{12\cdots n}^{{\rm U}_i},\label{totaluncorrelated}\\
&V_i^{\rm TC}=V_i^{\rm C}+\sum_{j\neq i}^nV_{ij}^{{\rm C}_i}+\cdots+V_{12\cdots n}^{{\rm C}_i},\label{totalcorrelated}\\
&V_i^{\rm TUC}=V_i^{\rm UC}+\sum_{j\neq i}^nV_{ij}^{{\rm UC}_i}+\cdots+V_{12\cdots n}^{{\rm UC}_i}.\label{totalun-correlated}
\end{align}
The sensitivity (or importance) measures are then spontaneously determined by
\begin{align}
&s_i^{\rm U}=\frac{V_i^{\rm U}}{V(y)}, \qquad
s_{i}^{\rm C}=\frac{V_{i}^{\rm C}}{V(y)},\qquad
s_{i}^{\rm UC}=\frac{V_{i}^{\rm UC}}{V(y)},\nonumber\\
&s_i^{\rm TU}=\frac{V_i^{\rm TU}}{V(y)},\qquad
s_i^{\rm TC}=\frac{V_i^{\rm TC}}{V(y)}, \qquad
s_i^{\rm TUC}=\frac{V_i^{\rm TUC}}{V(y)}. \label{correaltedsensitivitymeasures}
\end{align}
First three measures are called the main sensitivity indices which, separately, denote the importance of the independent section, correlated section and their coupling effect for $x_i$ before considering the interaction effects between $x_i$ and the remaining inputs. Last three measures are called the total sensitivity indices which, similarly, denote the importance of the independent section, correlated section and their coupling effect for $x_i$, by regarding the interaction effects of $x_i$ with the remaining inputs.

\section{Numerical examples and a practical application}
In this section, analytic polynomial models, including one purely additive and three nonlinear ones, are taken as examples to illustrate the effectiveness and validation of our established analytic framework. A practical application of the method is also proposed to the uncertainty and sensitivity analysis of a deterministic HIV model. Ten associated parameters are then ranked according to their importance in establishing the uncertainty of the basic reproduction number $R_0$.

\subsection{Additive linear model}
In the first example a purely additive model of the form as follows is investigated:
\begin{equation}
\label{model1}
y=2x_1+x_2+x_3,
\end{equation}
where $(x_1, x_2, x_3)\sim N(\bm{\mu},\Sigma)$ with mean vector $\bm{\mu}=(0, 0, 0)$ and covariance matrix
\begin{equation}
\Sigma=
\left(
  \begin{array}{ccc}
  1         & \rho_{12} & 2\rho_{13} \\
  \rho_{12} & 1         & 2\rho_{23} \\
  2\rho_{13}&2\rho_{23} & 4 \\
  \end{array}
\right).
\end{equation}

By the use of Eq. (\ref{vcorrelated}), the exact expression of the total variance $V(y)$ of model response is obtained as
\begin{equation}
V(y)=9+4\rho_{12}+8\rho_{13}+4\rho_{23},
\end{equation}
which is constituted of the fractional contributions of different dimensionality, including
\begin{equation}
V_1=4, \qquad V_2=1, \qquad V_3=4,\qquad V_{12}=4\rho_{12}, \qquad V_{13}=8\rho_{13}, \qquad
V_{23}=4\rho_{23}, \qquad V_{123}=0.\label{test1}
\end{equation}
Vanishing nonlinear problems in the functional form of the model under discussion suggest nonexistent coupling effect but only existent independent and correlated ones that are calculated as
\begin{align}
&V_i^{\rm U}=c_i^2V_i, \qquad V_i^{\rm C}=(1-c_i^2)V_i, \qquad
V_{12}^{{\rm U}_j}=V_{13}^{{\rm U}_k}=V_{23}^{{\rm U}_l}=V_{123}^{{\rm U}_i}=0, \nonumber \\
&V_{12}^{{\rm C}_j}=V_{12}, \qquad V_{13}^{{\rm C}_k}=V_{13}, \qquad V_{23}^{{\rm C}_l}=V_{23}, \qquad
V_{123}^{{\rm C}_i}=0,\label{test1thirdorder}
\end{align}
where $i\in \{1, 2, 3\}$, $j\in \{1,2\}$, $k\in \{1,3\}$, $l\in \{2,3\}$, and $c_i$, specifying the independence of variable $x_i$, is determined via Eq. (\ref{coefficientuncorrelated}) as
\begin{align}
c_1&=(1-\rho_{23}^2)^{-1/2}(1-\rho_{12}^2-\rho_{13}^2-\rho_{23}^2+2\rho_{12}\rho_{13}\rho_{23})^{1/2},\\
c_2&=(1-\rho_{13}^2)^{-1/2}(1-\rho_{12}^2-\rho_{13}^2-\rho_{23}^2+2\rho_{12}\rho_{13}\rho_{23})^{1/2},\\
c_3&=(1-\rho_{12}^2)^{-1/2}(1-\rho_{12}^2-\rho_{13}^2-\rho_{23}^2+2\rho_{12}\rho_{13}\rho_{23})^{1/2},
\end{align}
in which $\{\rho_{12}^2, \rho_{13}^2, \rho_{23}^2\}\neq 1$.

The underlying sensitivity measures are provided in Table \ref{t1} under considering different correlations between input variables. Results indicate vanishing sensitivity indices associated with the coupling between independent and correlated sections contained in each input variable. In the absence of correlated input variables ($\rho=0$), the main sensitivity indices sum up to one. By introducing input correlations, however, this summation  could be smaller than one (with positive correlations) or larger than one (with negative correlations), contrary to the sum of the total sensitivity indices. Negative sensitivity indices explain negative partial variance contributions produced by the negative input correlation.

\begin{table}
\centering
\caption{\label{t1} Uncertainty and sensitivity analysis results for linear additive model by assuming different correlations between input variables.}
{\footnotesize
\begin{tabular}{@{}*{11}{c}}
  \hline\hline
 $\rho$  & $V(y)$ & $x$ & $s_i$ & $s_i^{\rm U}$ & $s_i^{\rm C}$ & $s_i^{\rm UC}$ & $s_{Ti}$ & $s_{i}^{\rm TU}$ & $s_{i}^{\rm TC}$ & $s_{i}^{\rm TUC}$ \\
\hline

&  & $x_1$ & 0.444 & 0.444 & 0.0 & 0.0 & 0.444 & 0.444 & 0.0 & 0.0 \\
$\rho=0$  & 9 & $x_2$ & 0.111 & 0.111 & 0.0 & 0.0 & 0.111 & 0.111 & 0.0 & 0.0\\
& & $x_3$ & 0.444 & 0.444 & 0.0 & 0.0 & 0.444 & 0.444 & 0.0 & 0.0\\
\hline

 & & $x_1$ & 0.328 & 0.118 & 0.210 & 0.0 & 0.590 & 0.118 & 0.472 & 0.0 \\
$\rho_{12}=0.8$ & 12.2  & $x_2$ & 0.082 & 0.030 & 0.052 & 0.0 & 0.344 & 0.030 & 0.314 & 0.0\\
& & $x_3$ & 0.328 & 0.328 & 0.0 & 0.0 & 0.328 & 0.328 & 0.0 & 0.0\\
\hline

 &  & $x_1$ & 0.690 & 0.248 & 0.442 & 0.0 & 0.138 & 0.248 & -0.110 & 0.0 \\
$\rho_{12}=-0.8$ & 5.8 & $x_2$ & 0.172 & 0.062 & 0.110 & 0.0 & -0.379 & 0.062 & -0.441 & 0.0\\
& & $x_3$ & 0.690 & 0.690 & 0.0 & 0.0 & 0.690 & 0.690 & 0.0 & 0.0\\
\hline

$\rho_{12}=0.8$ &  & $x_1$ & 0.225 & 0.072 & 0.152 & 0.0 & 0.629 & 0.072 & 0.557 & 0.0\\
$\rho_{13}=0.5$ & 17.8 & $x_2$ & 0.056 & 0.020 & 0.036 & 0.0 & 0.326 & 0.020 & 0.306 & 0.0\\
$\rho_{23}=0.4$ & & $x_3$ & 0.225 & 0.169 & 0.056 & 0.0 & 0.539 & 0.169 & 0.371 & 0.0\\
\hline\hline
\end{tabular}
}
\end{table}

\subsection{Nonlinear models}
\subsubsection{Trivariate model}
In the second example a nonlinear model dependent upon three input variables is considered. It contains linear, quadratic, and interaction terms£º
\begin{equation}
\label{testcase2}
y=2x_1+x_2^2+4x_1^2x_2+x_1x_3,
\end{equation}
where $(x_1, x_2, x_3)\sim N(\bm{\mu},\Sigma)$ with mean vector $\bm{\mu}=(0, 0, 0)$ and covariance matrix
\begin{equation}
\label{testcase2covariancematrix}
\Sigma=
\left(
  \begin{array}{ccc}
  1 & \rho_{12} & \rho_{13} \\
  \rho_{12} & 1 & \rho_{23} \\
  \rho_{13} & \rho_{23} & 1 \\
  \end{array}
\right).
\end{equation}
The total variance $V(y)$ of model response is similarly computed by the use of Eq. (\ref{vcorrelated}) as
\begin{equation}
V(y)=55+48\rho_{12}+2\rho_{12}\rho_{23}+192\rho_{12}^2+\rho_{13}^2,
\end{equation}
which is generated by the partial variance contributions involving
\begin{align}
&V_1=4,\qquad V_2=2,\qquad V_3=V_{23}=0,\nonumber\\
&V_{12}=48(1+4\rho_{12}^2+\rho_{12}),\qquad
V_{13}=1+\rho_{13}^2,\qquad V_{123}=2\rho_{12}\rho_{23}.\label{x123}
\end{align}
The independent, correlated, and coupling effects divided from the main variance contributions $V_1$ and $V_2$ are stated as
\begin{align}
&V_1^{\rm U}=4c_1^2, \qquad V_1^{\rm C}=4(1-c_1^2), \qquad V_1^{\rm UC}=0, \nonumber\\
&V_2^{\rm U}=2c_2^4,\qquad V_2^{\rm C}=2(1-c_2^2)^2,\qquad V_2^{\rm UC}=4c_2^2(1-c_2^2)\label{testcase2maineffect}.
\end{align}
Regarding the existent higher order partial variance contributions, we have
\begin{align}
&V_{12}^{{\rm U}_1}=48c_1^4,\qquad V_{12}^{{\rm U}_2}=48c_2^2, \qquad V_{13}^{{\rm U}_i}=c_i^2,\qquad
V_{123}^{{\rm U}_j}=V_{123}^{{\rm UC}_j}=0\nonumber \\
&V_{12}^{{\rm C}_1}=48(1-c_1^2)(1-c_1^2+\rho_{12}+4\rho_{12}^2),\qquad
V_{12}^{{\rm UC}_1}=48(2+\rho_{12}+4\rho_{12}^2-2c_1^2)c_1^2,\nonumber\\
&V_{12}^{{\rm C}_2}=48(1+4\rho_{12}^2+\rho_{12}-c_2^2), \qquad V_{12}^{{\rm UC}_2}=0, \qquad V_{13}^{{\rm C}_i}=1+\rho_{13}^2-c_i^2\qquad
V_{123}^{{\rm C}_j}=2\rho_{12}\rho_{23},
\end{align}
where $i\in \{1, 3\}$, $j \in \{1,2,3\}$ and $c_i$ is determined with Eq. (\ref{coefficientuncorrelated}). A detailed calculation for the above items are presented in Appendix \ref{appendixtestcase2}. In Table \ref{t2}, the corresponding analysis results are listed, suggesting a dominated influence of the interaction effect between $x_1$ and $x_2$ for the case without correlated input variables. In the presence of input correlations, the independent, correlated, and coupling variance contributions produced by $x_1$ and $x_2$ are all significant for establishing the uncertainty of model response.

\begin{table}[t]
\centering
\caption{\label{t2} Analytic results for uncertainty and sensitivity analysis of the first nonlinear model with different input correlations.}
{\footnotesize
\begin{tabular}{@{}*{11}{l}}
  \hline\hline
 $\rho$  & $V(y)$ & $x$ & $s_i$ & $s_i^{\rm U}$ & $s_i^{\rm C}$ & $s_i^{\rm UC}$ & $s_{Ti}$ & $s_{i}^{\rm TU}$ & $s_{i}^{\rm TC}$ & $s_i^{\rm TUC}$ \\
\hline
 &  & $x_1$ & 0.073 & 0.073 & 0.0 & 0.0 & 0.964 & 0.964 & 0.0 & 0.0 \\
$\rho=0$ & 55 & $x_2$ & 0.036 & 0.036 & 0.0 & 0.0 & 0.909 & 0.909 & 0.0 & 0.0\\
& & $x_3$ & 0.0 & 0.0 & 0.0 & 0.0 & 0.018 & 0.018 & 0.0 & 0.0\\
\hline

&  & $x_1$ & 0.031 & 0.024 & 0.007 & 0.0 & 0.984 & 0.242 & 0.175 & 0.567 \\
$\rho_{12}=0.5$ & 127 & $x_2$ & 0.016 & 0.009 & 0.001 & 0.006 & 0.960 & 0.292 & 0.662 & 0.006\\
& & $x_3$ & 0.0 & 0.0 & 0.0 & 0.0 & 0.008 & 0.008 & 0.0 & 0.0\\
\hline

$\rho_{12}=-0.5$ &  & $x_1$ & 0.050 & 0.020 & 0.03 & 0.0 & 0.975 & 0.117 & 0.452 & 0.406\\
 $\rho_{13}=0.6$ & 79.36 & $x_2$ & 0.025 & 0.009 & 0.004 & 0.012 & 0.932 & 0.382 & 0.538 & 0.012\\
& & $x_3$ & 0.0 & 0.0 & 0.0 & 0.0 & 0.017 & 0.007 & 0.010 & 0.0\\
\hline

$\rho_{12}=0.4$ &  & $x_1$ & 0.038 & 0.028 & 0.01 & 0.0 & 0.981 & 0.291 & 0.166 & 0.524 \\
$\rho_{13}=0.5$ & 105.81 & $x_2$ & 0.019 & 0.002 & 0.008 & 0.009 & 0.950 & 0.167 & 0.774 & 0.009\\
$\rho_{23}=0.8$ & & $x_3$ & 0.0 & 0.0 & 0.0 & 0.0 & 0.018 & 0.003 & 0.015 & 0.0\\
\hline\hline
\end{tabular}
}
\end{table}

\subsubsection{Fourvariate model}

Another nonlinear model is designed based on four input variables as
\begin{equation}
y=x_1x_3+x_2x_4,
\end{equation}
where $(x_1, x_2, x_3, x_4)\sim N(\bm{\mu},\Sigma)$ with mean vector $\bm{\mu}=(1, 2, 2, 1)$ and covariance matrix
\begin{equation}
\label{testcase3covariance}
\Sigma=
\left(
\begin{array}{cccc}
  1 & \rho_{12} & \rho_{13} & \rho_{14} \\
  \rho_{12} & 1 & \rho_{23} & \rho_{24} \\
  \rho_{13} & \rho_{23} & 1 & \rho_{34}\\
  \rho_{14} & \rho_{24} & \rho_{34} & 1\\
\end{array}
\right)
.
\end{equation}
The total variance of model response is obtained by employing Eq. (\ref{vcorrelated}) as
\begin{equation}
V(y)=12+4(\rho_{12}+\rho_{13}+2\rho_{14}+\rho_{24}+\rho_{34})+2\rho_{23}+\rho_{13}^2+\rho_{24}^2+2(\rho_{12}\rho_{34}+\rho_{14}\rho_{23})\label{test3},
\end{equation}
which is constituted of
\begin{align}
&V_1=4, \qquad V_2=1, \qquad V_3=1, \qquad V_4=4,\nonumber\\
&V_{12}=4\rho_{12}, \qquad V_{13}=1+4\rho_{13}+\rho_{13}^2, \qquad V_{14}=8\rho_{14}, \qquad V_{23}=2\rho_{23}, \nonumber\\
&V_{24}=1+4\rho_{24}+\rho_{24}^2, \qquad V_{34}=4\rho_{34},\qquad V_{1234}=2(\rho_{12}\rho_{34}+\rho_{14}\rho_{23}).
\end{align}
In the evaluation of $V_{1234}$, the first-order covariance ${\rm cov}(x_1, x_2, x_3, x_4)$ of four correlated variables is involved. Its detailed derivation is presented in Appendix \ref{appendixtestcase3}. The form of model function (only involves the linear problem of each input) suggests the vanishing coupling effects in all partial variance contributions but existent correlated and independent ones. We get
\begin{align}
&V_{i}^{\rm U}=c_i^2V_i, \qquad V_{13}^{{\rm U}_j}=c_j^2, \qquad V_{24}^{{\rm U}_k}=c_k^2, \nonumber\\
&V_{i}^{\rm C}=(1-c_i^2)V_i, \qquad V_{13}^{{\rm C}_j}=V_{13}-V_{13}^{{\rm U}_j}, \qquad V_{24}^{{\rm C}_k}=V_{24}-V_{24}^{{\rm U}_k}, \label{test3seconduncorrelated}
\end{align}
where $i\in \{1, 2, 3, 4\}$, $j\in\{1, 3\}$ and $k\in\{2, 4\}$. Other partial variance contributions $V_{12}$, $V_{14}$, $V_{23}$, $V_{34}$ and $V_{1234}$ are all contributed by the correlations of involved parameters. The coefficient $c_i$ is determined with Eq. (\ref{coefficientuncorrelated}). Table \ref{t3} lists the analytic values of the underlying sensitivity indices. Data show a vanishing coupling effect between the correlated and independent sections of each individual variable. This because nonlinear problems of single variables are absent in the form of model function.

\begin{table}
\centering
\caption{\label{t3} Analytic values of uncertainty and sensitivity analysis for the second nonlinear model by assuming uncorrelated and correlated inputs.}
{\footnotesize
\begin{tabular}{@{}*{11}{c}}
  \hline\hline
 $\rho$  & $V(y)$ & $x$ & $s_i$ & $s_i^{\rm U}$ & $s_i^{\rm C}$ & $s_i^{\rm UC}$ & $s_{Ti}$ & $s_{i}^{\rm TU}$ & $s_{i}^{\rm TC}$ & $s_i^{\rm TUC}$\\
\hline
&  & $x_1$ & 0.333 & 0.333 & 0.0 & 0.0 & 0.417 & 0.417 & 0.0 & 0.0\\
$\rho=0$ & 12 & $x_2$ & 0.083 & 0.083 & 0.0 & 0.0 & 0.167 & 0.167 & 0.0 & 0.0\\
& & $x_3$ & 0.083 & 0.083 & 0.0 & 0.0 & 0.167 & 0.167 & 0.0 & 0.0\\
& & $x_4$ & 0.333 & 0.333 & 0.0 & 0.0 & 0.417 & 0.417 & 0.0 & 0.0\\
\hline

$\rho_{13}=0.5$ &  & $x_1$ & 0.221 & 0.166 & 0.055 & 0.0 & 0.401 & 0.207 & 0.194 & 0.0 \\
$\rho_{24}=0.8$& 18.09 & $x_2$ & 0.055 & 0.020 & 0.035 & 0.0 & 0.323 & 0.040 & 0.283 & 0.0\\
& & $x_3$ & 0.055 & 0.041 & 0.014 & 0.0 & 0.235 & 0.083 & 0.152 & 0.0\\
& & $x_4$ & 0.221 & 0.080 & 0.141 & 0.0 & 0.489 & 0.100 & 0.389 & 0.0\\
\hline

$\rho_{12}=-0.5$ &  & $x_1$ & 0.251 & 0.058 & 0.193 & 0.0 & 0.561 & 0.072 & 0.489 & 0.0 \\
$\rho_{13}=0.6$ & 15.96 & $x_2$ & 0.063 & 0.030 & 0.033 & 0.0 & 0.0 & 0.060 & -0.060 & 0.0\\
$\rho_{14}=0.4$ & & $x_3$ & 0.063 & 0.024 & 0.039 & 0.0 & 0.298 & 0.049 & 0.249 & 0.0\\
& & $x_4$ & 0.251 & 0.148 & 0.103 & 0.0 & 0.514 & 0.185 & 0.329 & 0.0\\
\hline

$\rho_{12}=-0.5$, $\rho_{13}=-0.4$ &  & $x_1$ & 0.350 & 0.196 & 0.154 & 0.0 & 0.322 & 0.245 & 0.077 & 0.0 \\
$\rho_{14}=0.2$, $\rho_{23}=0.3$   &    11.44   & $x_2$ & 0.087 & 0.034 & 0.053 & 0.0 & 0.252 & 0.069 & 0.183 & 0.0\\
$\rho_{24}=0.4$, $\rho_{34}=0.4$   &       & $x_3$ & 0.087 & 0.051 & 0.036 & 0.0 & 0.007 & 0.103 & -0.096 & 0.0\\
& & $x_4$ & 0.350 & 0.156 & 0.194 & 0.0 & 0.636 & 0.195 & 0.441 & 0.0 \\
\hline\hline
\end{tabular}
}
\end{table}

\subsubsection{Ishigami function}
The Ishigami function \cite{Ishigami1990} has been extensively used as
a benchmark for sensitivity analysis \cite{Morio2011,Lira2016}. Its functional form was defined as
\begin{equation}
y=\sin(x_1)+7\sin^2(x_2)+0.1x_3^4\sin(x_1)\label{ishigamifunction},
\end{equation}
where all input variables are uniformly distributed in the interval $[-\pi, \pi]$. The presence of correlation between $x_2$ and each of the rest does not influence the total variance of model response owing to zero partial variance contributions associated with the interaction between $x_2$ and the rest. Consequently, we just consider here the correlation between $x_1$ and $x_3$. The results of analytic analysis are listed in Table \ref{t4} by assuming independent and correlated input variables. Two formulating strategies are considered in the presence of correlation: $x_1$ is formulated on the basis of $x_3$ and vice versa. They are non-equivalent for the uncertainty and sensitivity analysis of the model under discussion as $x_1$ and $x_3$ are uniformly distributed.

In the first case $x_1$ is formulated on the basis of $x_3$ as
\begin{equation}
x_1=\rho_{13}\frac{\sigma_1}{\sigma_3}x_3+\sqrt{1-\rho_{13}^2}r_1,
\end{equation}
where the newly introduced random variable $r_1$ holds the same distribution with $x_1$. In the form of Ishigami function, the contribution of $x_1$ to the variance of model response is embodied by $\sin$ function which explains the marginal effect of $x_1$ is totally produced by the coupling between its independent and correlated sections. Sensitivity measures for this case state a strong positive variance contribution produced by the interaction effect between $x_3$ and the correlated part of $x_1$, as well as a very strong negative variance contribution caused by the interaction term involving $x_3$ and both the correlated and independent sections of $x_1$.

For the second case, we generate $x_3$ on the basis of $x_1$ as
\begin{equation}
x_3=\rho_{13}\frac{\sigma_3}{\sigma_1}x_1+\sqrt{1-\rho_{13}^2}r_3,
\end{equation}
where random variable $r_3$ follows the same distribution with $x_3$. Zero mean of $x_1$ leads to the nonexistence of sensitivity measures associated with the main effect of $x_3$. A dominated contribution to the variance of model response is produced by the interaction effect between $x_1$ and the coupling of independence with correlated section of $x_3$.

Sensitivity indices of correlated and independent sections of $x_3$ are not indicated in the first case because $x_3$ is not decomposed into independent and correlated parts but only considered as a whole variable, analogous to the second case. The derivation process of partial variance contributions of different orders is presented in detail in Appendix \ref{ishigamifunction2} for both cases.

\begin{table}
\centering
\caption{\label{t4} Analytic values of uncertainty and sensitivity analysis for Ishigami function by assuming uncorrelated and correlated input variables. In case 1 $x_1$ is generated based on $x_3$, contrary to the second case where $x_3$ is generated based on $x_1$.}
{\footnotesize
\begin{tabular}{@{}*{11}{c}}
  \hline\hline
 $\rho$  & $V(y)$ & $x$ & $s_i$ & $s_i^{\rm U}$ & $s_i^{\rm C}$ & $s_i^{\rm UC}$ & $s_{Ti}$ & $s_{i}^{\rm TU}$ & $s_{i}^{\rm TC}$ & $s_i^{\rm TUC}$\\
\hline

$\rho_{13}=0$ &  & $x_1$ & 0.036 & 0.036 & 0.0 & 0.0 & 0.557 & 0.557 & 0.0 & 0.0\\
& 13.845 & $x_2$ & 0.442 & 0.442 & 0.0 & 0.0 & 0.442 & 0.442 & 0.0 & 0.0\\
& & $x_3$ & 0.0 & 0.0 & 0.0 & 0.0 & 0.527 & 0.527 & 0.0 & 0.0\\
\hline

$\rho_{13}=0.5$ &  & $x_1$ & 0.039 & 0.0 & 0.0 & 0.039 & 0.528 & 0.633 & 1.048 & -1.153 \\
(case 1)& 12.971 & $x_2$ & 0.472 & 0.472 & 0.0 & 0.0 & 0.472 & 0.472 & 0.0 & 0.0\\
& & $x_3$ & 0.0 & --- & --- & --- & 0.489 & --- & --- & --- \\
\hline

$\rho_{13}=0.5$ &  & $x_1$ & 0.026 & --- & --- & --- & 0.679 & --- & --- & --- \\
(case two)& 19.110 & $x_2$ & 0.321 & 0.321 & 0.0 & 0.0 & 0.321 & 0.321 & 0.0 & 0.0\\
& & $x_3$ & 0.0 & 0.0 & 0.0 & 0.0 & 0.653 & 0.145 & 0.004 & 0.505\\
\hline\hline
\end{tabular}
}
\end{table}

\subsection{HIV model}
The basic reproduction number, denoted as $R_0$, is arguably the most important quantity in infectious disease epidemiology because it helps determine whether or not an infectious disease can spread through a population \cite{Fraser2009,Holme2015}. $R_0$ is defined as the average number of new cases of an infection caused by one typical infected individual, in a population consisting of susceptibles only \cite{anderson1992,diekmann1990,Diekmann2010}. The first application of this metric in epidemiology was introduced by George MacDonald in 1952, who designed agents-based models of the spread of malaria \cite{Macdonald1957}. Generally, the larger the value of $R_0$, the harder it is to control the spreading of an epidemic. Typically, when $R_0<1$, the disease free equilibrium is locally asymptotically stable and the epidemic will die out in the long run, whereas if $R_0>1$, it is unstable and the epidemic will invade the population \cite{Van2002}.

Consider a deterministic model of HIV-1 with vertical transmission (from an HIV-infected mother to her child) which was discussed in Ref. \cite{Usman2016}. The basic reproduction number $R_0$ is represented by
\begin{equation}
\label{hivmodel}
R_0=\frac{\beta_0(1-\gamma)\theta_d^2+\beta_1n_1Q_0(\theta_d-\kappa)+\beta_2n_2\alpha Q_0+(1-\gamma)(\kappa+\alpha)\beta_0\theta_d}{\theta_d(\theta_d+\kappa)(\theta_d+\alpha)}.
\end{equation}
Description and baseline values of parameters included in the above expression are presented in Table \ref{t5}.

\begin{table}
\centering
\caption{\label{t5} Description and baseline values of parameters for HIV/AIDS model, see Refs. \cite{Usman2016,Mukandavire2006,Safiel2012}.}
{\footnotesize
\begin{tabular}{lll}
  \hline\hline
Parameter & Symbol & Baseline value \\
\hline
Recruitment rate & $Q_0$ & 0.029\\
Birth rate of infective & $\beta_0$ & 0.03 \\
Fraction of susceptible newborn & \multirow{2}{*}{$\gamma$} & \multirow{2}{*}{0.4} \\
from infective class & & \\
Contact rate of susceptible with & \multirow{2}{*}{$\beta_1$} & \multirow{2}{*}{0.2} \\
asymptomatic infective & &\\
Contact rate of susceptible with & \multirow{2}{*}{$\beta_2$} & \multirow{2}{*}{0.08} \\
symptomatic infective & &\\
Number of sexual partners of susceptible & \multirow{2}{*}{$n_1$} & \multirow{2}{*}{2.0} \\
with asymptomatic infective & & \\
Number of sexual partners of susceptible & \multirow{2}{*}{$n_2$} & \multirow{2}{*}{2.0} \\
with symptomatic infective & & \\
Natural death rate & $\theta_d$ & 0.02 \\
Removal rate to symptomatic class & $\alpha$ & 0.6 \\
Rate of development to AIDS & $\kappa$ & 0.1 \\
\hline\hline
\end{tabular}
}
\end{table}

\begin{figure}
\centering
\includegraphics[width=6cm]{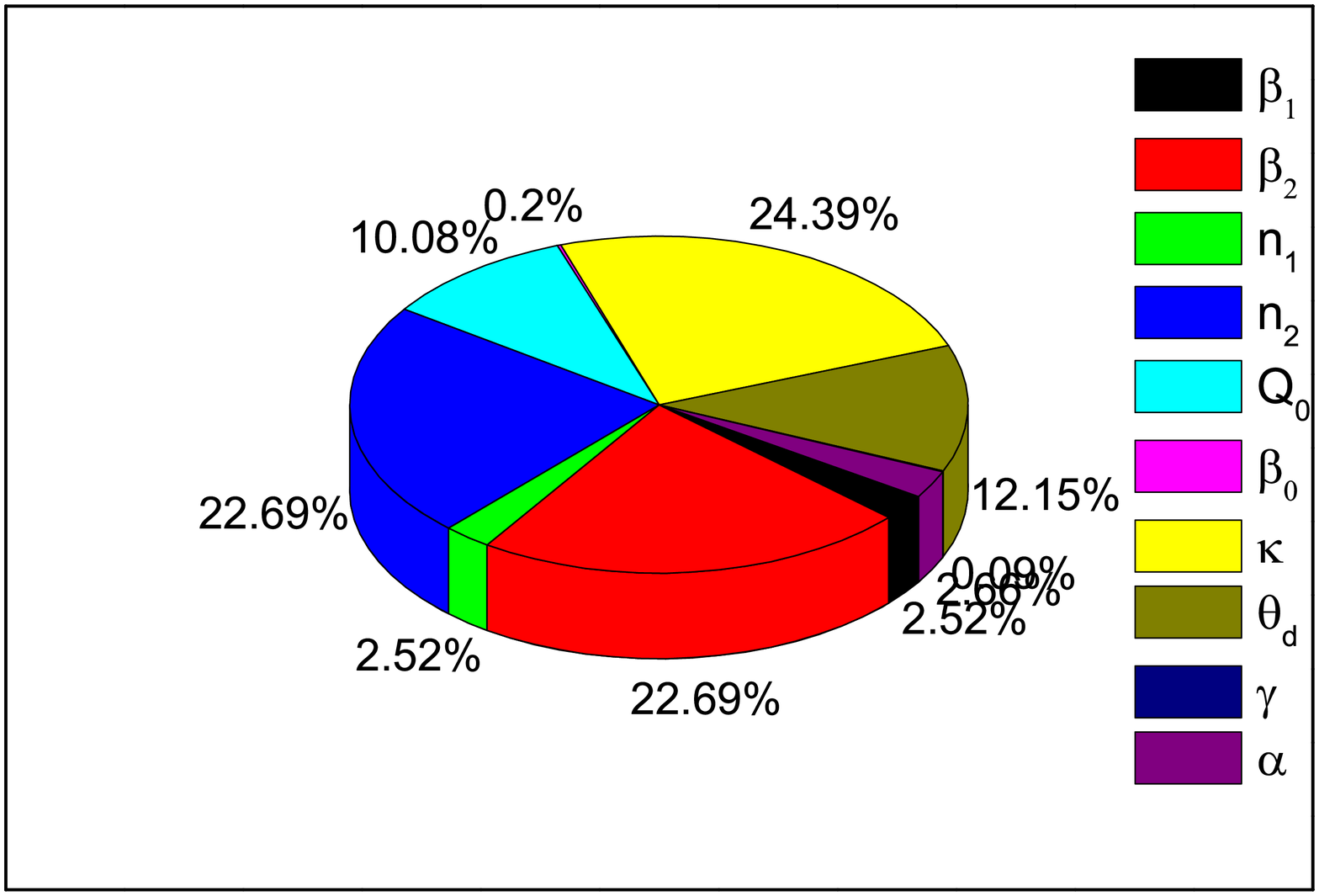}
\includegraphics[width=6cm]{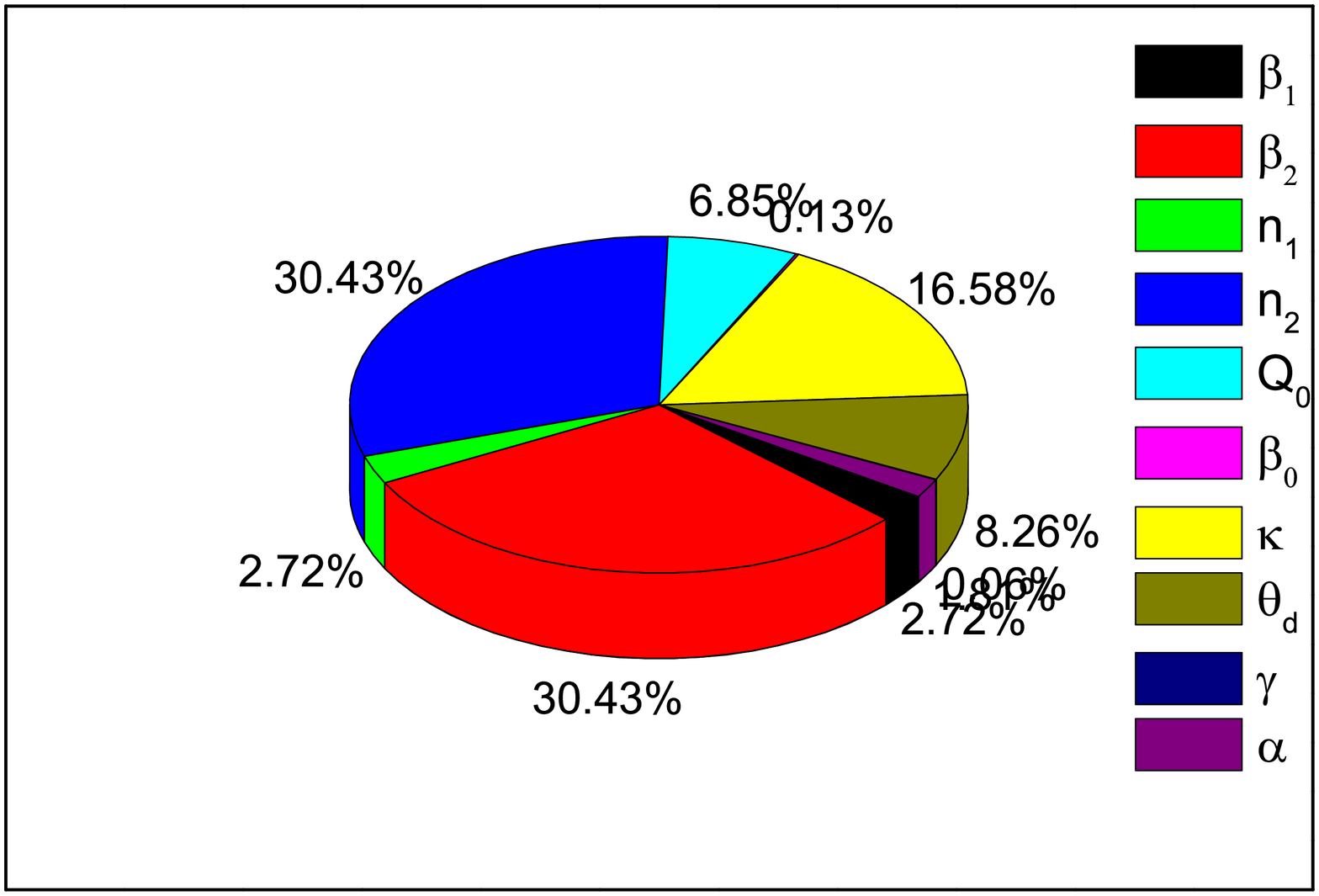}
\caption{(Color online) Parameters are ranked according to their importance (sensitivity indices $s_i$) in establishing the uncertainty of the basic reproduction number $R_0$ for HIV-1 model with vertical transmission. Left panel is for independent situation, and right one for the correlated one with $\rho_{\beta_1n_1}=0.3$ and $\rho_{\beta_2n_2}=0.5$.}
\label{hivsensitivity}
\end{figure}

To identify the importance of individual parameters in establishing the uncertainty of $R_0$, each parameter is artificially increased and decreased 10\% of its baseline value. Furthermore, for simplicity, uncertainties of parameters are indicated by uniform distribution in their ranges of variation. The mathematical expectations and uncertainties of input parameters and output $R_0$ are presented in Table \ref{t6}. Regarding the uncertainty in $R_0$, both independent and correlated situations are discussed. The underlying sensitivity analysis results are displayed in Table \ref{t7}. In our analysis, the first-order partial variance contributions are considered only, which explain 99.6\% and 98.8\% of the exact uncertainty (indicated by the standard deviation) of $R_0$ for independent and correlated situations, respectively. In the presence of correlated parameters, the first-order contributions also contain the first-order interaction effects that are provided by input correlations. Sensitivity indices demonstrate that the top three parameters ranked according to their importance in establishing the uncertainty of $R_0$ are $\kappa$ (rate of development to AIDS), $\beta_2$ (contact rate of susceptible with symptomatic infective), and $n_2$ (number of sexual partners of susceptible with symptomatic infective) for both situations, see Fig. \ref{hivsensitivity}.

\begin{table}
\centering
\caption{\label{t6} Uncertainty determination for both input and output parameters. The uncertainty in $R_0$ presented here just contains the first-order partial variance contributions of individual input parameters. They explain 99.6\% and 98.8\% of the exact uncertainty of $R_0$ for independent and correlated situations, respectively.}
{\footnotesize
\begin{tabular}{cclll}
  \hline\hline
& parameter & $\mu$ (baseline value) & Range of variation & $\sigma$ (uncertainty) \\
\hline
\multirow{10}{*}{input} &
 $Q_0$ & 0.029 & [0.0261, 0.0319] & 0.002\\
& $\beta_0$ & 0.03 & [0.027, 0.033] & 0.002\\
& $\gamma$ & 0.4 & [0.36, 0.44] & 0.023\\
& $\beta_1$ & 0.2 & [0.18, 0.22] & 0.012\\
& $\beta_2$ & 0.08 & [0.072, 0.088] & 0.005\\
& $n_1$ & 2 & [1.8, 2.2] & 0.115\\
& $n_2$ & 2 & [1.8, 2.2] & 0.115\\
& $\theta_d$ & 0.02 & [0.018, 0.022] & 0.001\\
& $\alpha$ & 0.6 & [0.54, 0.66] & 0.035\\
& $\kappa$ & 0.1 & [0.09, 0.11] & 0.006\\
\hline
\multirow{2}{*}{output} &
\multirow{2}{*}{$R_0$} & 1.429 & -- & 0.227($\rho=0$) \\
\cline{3-5}
& & 1.432 & -- & 0.252 \\
& &       &    & ($\rho_{\beta_1n_1}=0.3$, $\rho_{\beta_2n_2}=0.5$)\\
\hline\hline
\end{tabular}
}
\end{table}

\begin{table}
\centering
\caption{\label{t7} Sensitivity analysis results of the basic reproduction number $R_0$ for a deterministic HIV-1 model with vertical transmission by assuming independent and correlated input parameters.}
{\footnotesize
\begin{tabular}{cc|ccc}
  \hline\hline
\multirow{2}{*}{\tabincell{c}{$x$}} & $\rho=0$ & \multicolumn{3}{|c}{$\rho_{\beta_1n_1}=0.3$, $\rho_{\beta_2n_2}=0.5$}\\
\cline{2-5}
& $s_i(=s_{i}^{\rm U})$ & $s_{i}$ & $s_{i}^{\rm U}$ & $s_{i}^{\rm C}$\\
\hline
$Q_0$ & 0.101 & 0.082 & 0.082 & 0\\
$\beta_0$ & 0.002 & 0.002 & 0.002 & 0\\
$\gamma$ & 0.001 &  0.001 & 0.001 & 0\\
$\beta_1$  & 0.025 & 0.032 & 0.018 & 0.014\\
$\beta_2$  & 0.227 & 0.363 & 0.138 & 0.225 \\
$n_1$ & 0.025 &0.032 & 0.018 & 0.014\\
$n_2$ & 0.227 & 0.363 & 0.138 & 0.225\\
$\theta_d$ & 0.122 & 0.098 & 0.098 & 0\\
$\alpha$ & 0.027 & 0.022 & 0.022 & 0\\
$\kappa$ & 0.244 & 0.197 & 0.197 & 0\\
\hline\hline
\end{tabular}
}
\end{table}

\begin{figure}
\centering
\includegraphics[width=6cm]{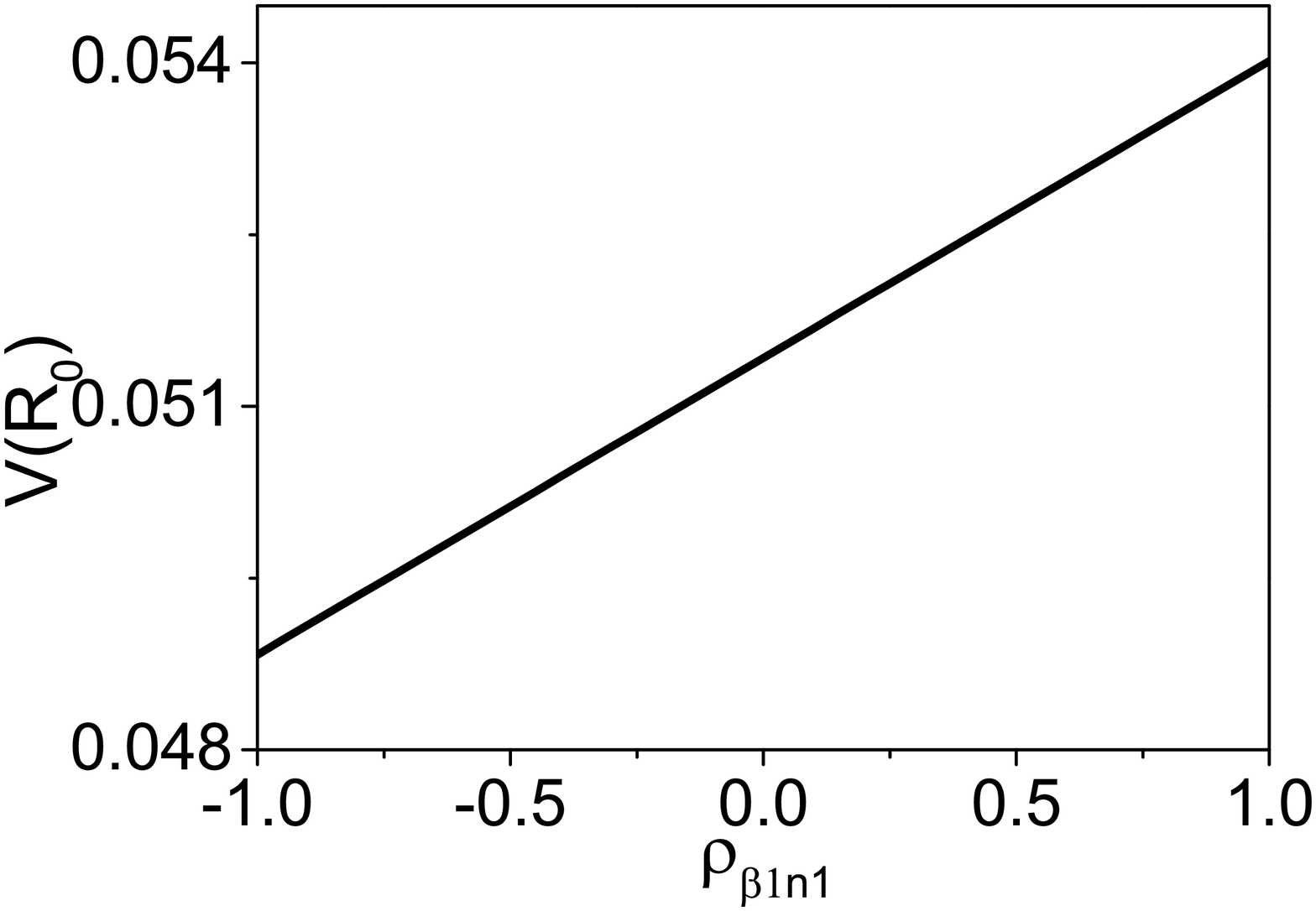}
\includegraphics[width=6cm]{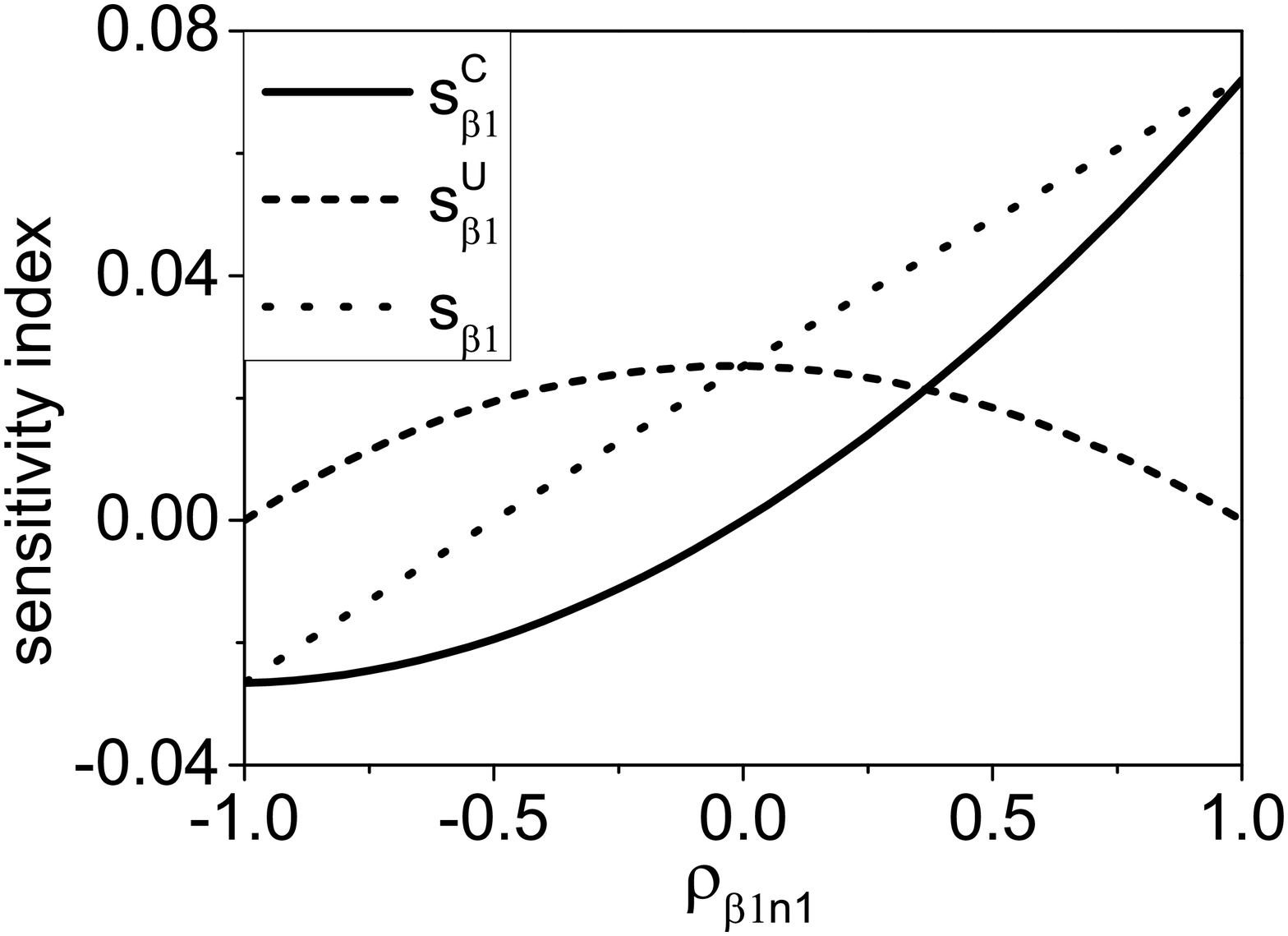}\\
(a) $\rho_{\beta_2n_2=0}$\\
\includegraphics[width=6cm]{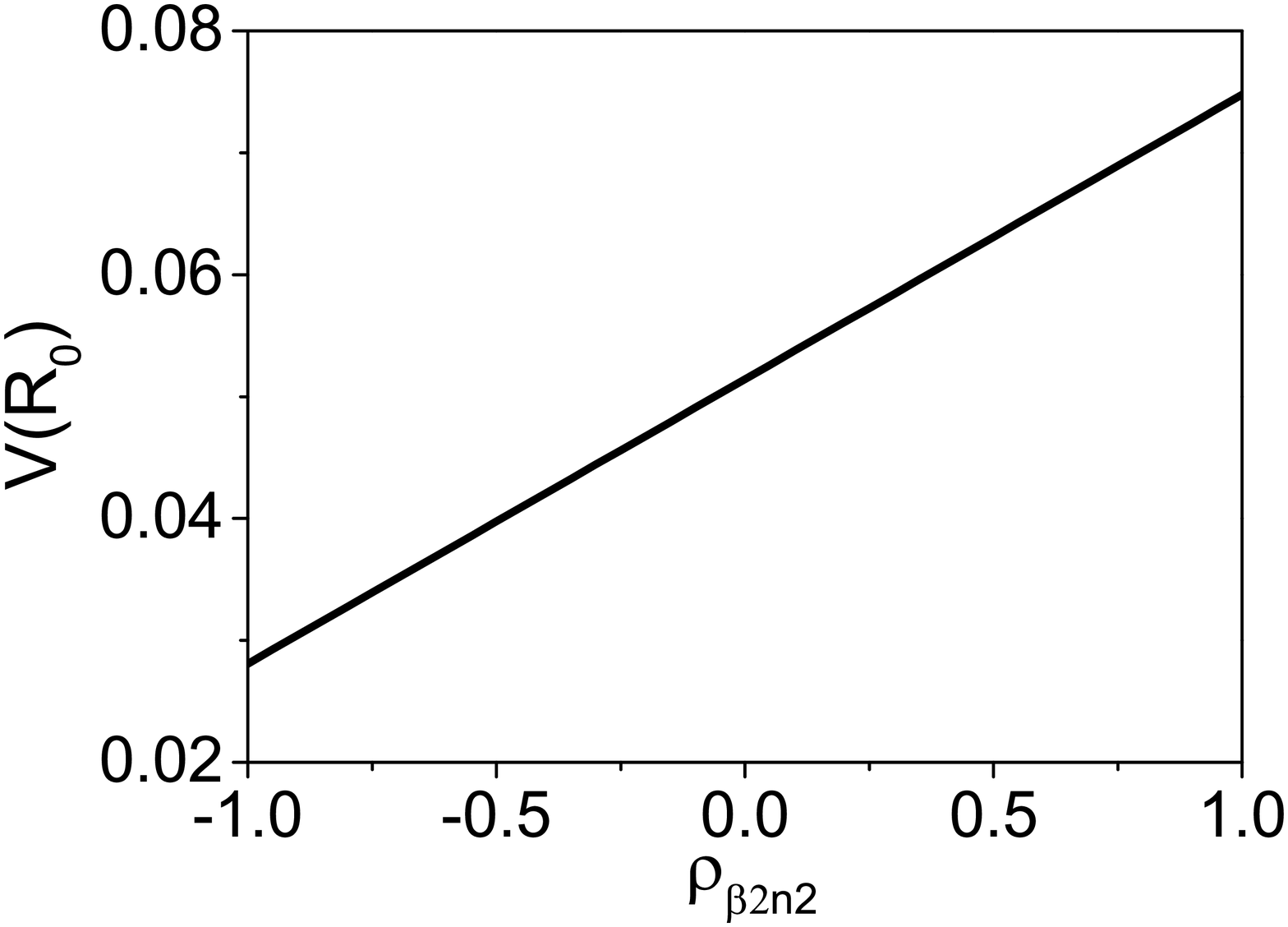}
\includegraphics[width=6cm]{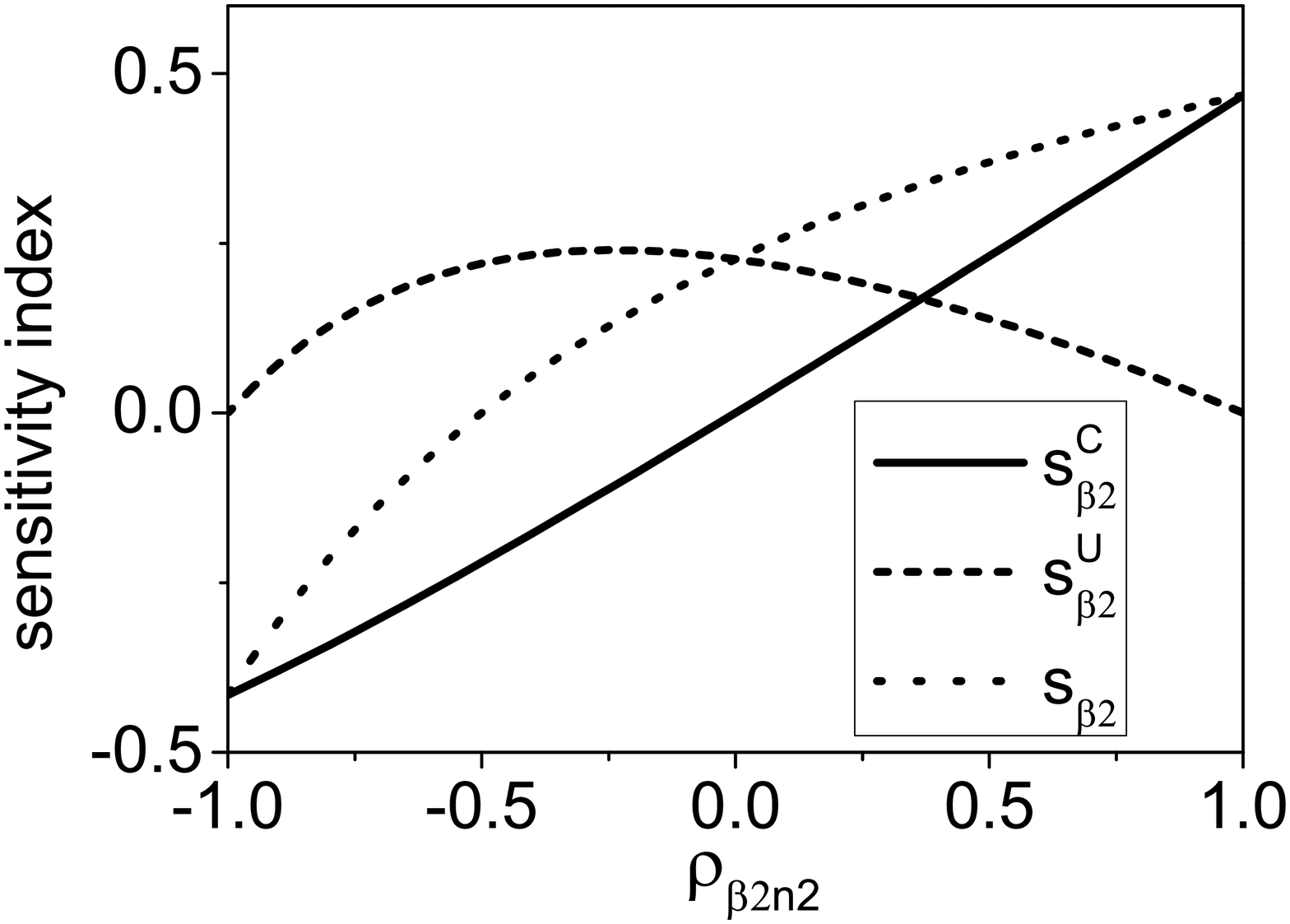}\\
(b) $\rho_{\beta_1n_1=0}$\\
\caption{The influence of correlations between $\beta_1$ and $n_1$, and between $\beta_2$ and $n_2$ on the uncertainty of $R_0$ and sensitivity indices of correlated parameters. The sensitivity indices for $n_1$ and $n_2$ are exactly the same as those for $\beta_1$ and $\beta_2$, respectively.}
\label{hivmodel2}
\end{figure}

The dependence of uncertainty in $R_0$ upon correlations between $\beta_1$ and $n_1$, and between $\beta_2$ and $n_2$ are showed in Fig. \ref{hivmodel2}. The corresponding distributions of sensitivity indices for correlated parameters are also presented. Distribution lines suggest that the correlation between $\beta_1$ and $n_1$ plays a fragile role in establishing the uncertainty of $R_0$. However, the correlation between $\beta_2$ and $n_2$ is non-negligible in determining $R_0$.

\section{Conclusions}
An analytic formula for variance propagation is extended to a general case with correlated input variables. It analytically evaluates the partial contributions produced by individual inputs alone, by their interactions and also by their correlations (if exist) to the variance of model response. Furthermore, the generation process of correlated variables is also analyzed. An arbitrary variable can be represented as the sum of independent and correlated sections. Universal expressions of the coefficients that specify the correlated and independent sections of a single variable are then derived by the use of linear correlation model. Based on the coefficients and analytic formula for variance propagation, it is straightforward to quantify the sensitivity of model response with respect to the input independence and correlations, as well as to their coupling effects. Four numerical examples have confirmed the effectiveness and applicability of our analytic framework. Finally, the analytic framework is applied to the uncertainty and sensitivity analysis of a deterministic HIV model with vertical transmission. Analysis results provide the importance of ten associated factors in determining the basic reproduction number $R_0$. This may help effectively decrease the spreading of HIV by controlling first three most important parameters including the rate of development to AIDS ($k$), the contact rate of susceptible with symptomatic infective ($\beta_2$), and the number of sexual partners of susceptible with symptomatic infective ($n_2$). Moreover, the correlation between $\beta_2$ and $n_2$, if exists, also provides a non-negligible effect on the uncertainty of the basic reproduction number.

\section*{Acknowledgements}
This work was supported by the Programme of Introducing Talents of Discipline to Universities under Grant No. B08033 and the fellowship from China Scholarship Council under Grant No. 201406770035. The authors are also grateful to Prof. Alain Bulou from Le Mans Universit\'e for his constant support and encouragement.

\appendix

\numberwithin{equation}{section}

\section{Central moments}\label{appendixcentral}
For a general uniformly distributed variable $x$ with standard deviation $\sigma$, the $k$th central moment is universally represented as
\begin{equation}
\label{centralmomentuniform}
M_k(x)=
\left\{
  \begin{array}{cc}
    \frac{3^{k/2}}{k+1}\sigma^k, &  k\quad {\rm is\quad even} \\
    0, &  k\quad {\rm is\quad odd}
  \end{array}
\right..
\end{equation}
If $x$ follows normal (Gaussian) distribution, the $k$th central moment is expressed by
\begin{equation}
\label{centralmomentnormal}
M_k(x)=
\left\{
  \begin{array}{cc}
    \sigma^k(k-1)!!, &  k\quad {\rm is\quad even} \\
    0, &  k\quad {\rm is\quad odd}
  \end{array}
\right.
,
\end{equation}
where $(\cdot)!!$ is the double factorial with $(-1)!!=1$. $k$ should be non-negative for both distributions.

\section{Generation of correlated variables}\label{appendixlinearcorrealtion}

\subsection{Two correlated variables}
Suppose $x_1$ is only correlated with $x_2$ in the input space with given correlation coefficient $\rho_{12}$. By using the linear correlation model, $x_1$ can be formulated on the basis of $x_2$ as
\begin{equation}
\label{twox12}
x_1=a_{12}x_2+c_1r_1,
\end{equation}
where $r_1$ is a random variable, holding the same variance with $x_1$. Employing the definition of correlation coefficient, we have
\begin{equation}
\begin{split}
\rho_{12}&=\frac{1}{\sigma_1\sigma_2}\int(x_1-\bar{x}_1)(x_2-\bar{x}_2)P(x_1, x_2){\rm d}x_1{\rm d}x_2\\
&=\frac{1}{\sigma_1\sigma_2}\int(a_{12}x_2+c_1r_1-a_{12}\bar{x}_2-c_1\bar{r}_1)(x_2-\bar{x}_2)P(x_2|x_1)P(r_1){\rm d}x_2{\rm d}r_1\\
&=\frac{\sigma_2}{\sigma_1}a_{12},
\end{split}
\end{equation}
which yields
\begin{equation}
a_{12}=\frac{\sigma_1}{\sigma_2}\rho_{12}.
\end{equation}
Furthermore, we also have
\begin{equation}
\begin{split}
\sigma_1^2&=\int(x_1-\bar{x}_1)^2P(x_1|x_2){\rm d}x_1\\
&=\int(a_{12}x_2+c_1r_1-a_{12}\bar{x}_2-c_1\bar{r}_1)^2P(x_2|x_1)P(r_1){\rm d}x_1{\rm d}r_1\\
&=a_{12}^2\sigma_2^2+c_1^2V(r_1).
\end{split}
\end{equation}
By substituting $\sigma_1^2=V(r_1)$, coefficient $c_1$ is obtained as
\begin{equation}
c_1=\sqrt{1-\rho_{12}^2}.
\end{equation}
As a consequence, $x_1$, if only correlated with $x_2$, can be formulated as
\begin{equation}
x_1=\frac{\sigma_1}{\sigma_2}\rho_{12}x_2+\sqrt{1-\rho_{12}^2}r_1.
\end{equation}

\subsection{Three correlated variables}
If $x_1$ is correlated with two variables, say $x_2$ and $x_3$, simultaneously, it can be similarly formulated by the use of linear correlation model as
\begin{equation}
x_1=a_{12}x_2+a_{13}x_3+c_1r_1.
\end{equation}
Recalling the correlation coefficient between $x_1$ and $x_2$, we get
\begin{equation}
\label{threerho12}
\begin{split}
\rho_{12}=&\frac{1}{\sigma_1\sigma_2}\int(x_1-\bar{x}_1)(x_2-\bar{x}_2)P(x_1, x_2|x_3){\rm d}x_1{\rm d}x_2\\
=&\frac{1}{\sigma_1\sigma_2}\int\left[a_{12}(x_2-\bar{x}_2)+a_{13}(x_3-\bar{x}_3)+c_1(r_1-\bar{r}_1)\right](x_2-\bar{x}_2)\\
&\times P(x_2, x_3|x_1)P(r_1){\rm d}x_2{\rm d}x_3{\rm d}r_1\\
=&\frac{\sigma_2}{\sigma_1}a_{12}+\frac{\sigma_3}{\sigma_1}a_{13}\rho_{23}.
\end{split}
\end{equation}
Analogously, the correlation coefficient between $x_1$ and $x_3$ can be expressed as
\begin{equation}
\begin{split}
\rho_{13}=&\frac{1}{\sigma_1\sigma_3}\int(x_1-\bar{x}_1)(x_3-\bar{x}_3)P(x_1, x_3|x_3){\rm d}x_1{\rm d}x_3\\
=&\frac{1}{\sigma_1\sigma_3}\int\left[a_{12}(x_2-\bar{x}_2)+a_{13}(x_3-\bar{x}_3)+c_1(r_1-\bar{r}_1)\right](x_3-\bar{x}_3)\\
&\times P(x_2, x_3|x_1)P(r_1){\rm d}x_2{\rm d}x_3{\rm d}r_1\\
=&\frac{\sigma_2}{\sigma_1}a_{12}\rho_{23}+\frac{\sigma_3}{\sigma_1}a_{13},
\end{split}
\end{equation}
which, together with Eq. (\ref{threerho12}), states
\begin{equation}
\label{threea}
a_{12}=\frac{\rho_{12}-\rho_{13}\rho_{23}}{1-\rho_{23}^2}\frac{\sigma_1}{\sigma_2}, \qquad a_{13}=\frac{\rho_{13}-\rho_{12}\rho_{23}}{1-\rho_{23}^2}\frac{\sigma_1}{\sigma_3}.
\end{equation}
$c_1$ can be determined by the definition of the variance of $x_1$, that is
\begin{equation}
\begin{split}
\sigma_1^2&=\int(x_1-\bar{x}_1)^2P(x_1|x_2, x_3){\rm d}x_1\\
&=\int\left[a_{12}(x_2-\bar{x}_2)+a_{13}(x_3-\bar{x}_3)+c_1(r_1-\bar{r}_1)\right]^2P(x_2, x_3|x_1)P(r_1){\rm d}x_2{\rm d}x_3{\rm d}r_1\\
&=a_{12}^2\sigma_2^2+a_{13}^2\sigma_3^2+2a_{12}a_{13}\sigma_2\sigma_3\rho_{23}+c_1^2V(r_1).
\end{split}
\end{equation}
Inserting the expressions of $a_{12}$ and $a_{13}$, and $\sigma_1^2=V(r_1)$, into the above equation, we obtain
\begin{equation}
\label{threec}
c_1=\sqrt{\frac{1-\rho_{12}^2-\rho_{13}^2-\rho_{23}^2+2\rho_{12}\rho_{13}\rho_{23}}{1-\rho_{23}^2}}.
\end{equation}
Accordingly, an arbitrary variable $x_1$ that is correlated with $x_2$ and $x_3$ at the same time can be generated with given pairwise correlation coefficients through equation
\begin{multline}
x_1=\frac{\rho_{12}-\rho_{13}\rho_{23}}{1-\rho_{23}^2}\frac{\sigma_1}{\sigma_2}x_2+\frac{\rho_{13}-\rho_{12}\rho_{23}}{1-\rho_{23}^2}\frac{\sigma_1}{\sigma_3}x_3
+\sqrt{\frac{1-\rho_{12}^2-\rho_{13}^2-\rho_{23}^2+2\rho_{12}\rho_{13}\rho_{23}}{1-\rho_{23}^2}}r_1.
\end{multline}

\subsection{Four correlated variables}
Consider a more complex situation where $x_1$ is correlated with three other variables, say $x_2$, $x_3$ and $x_4$, at the same time. As before, employing the linear correlation model, we can represent $x_1$ as
\begin{equation}
\label{fourx1}
x_1=a_{12}x_2+a_{13}x_3+a_{14}x_4+c_1r_1,
\end{equation}
in keeping $V(r_1)=\sigma_1^2$. Reviewing the definition of correlation coefficient, we have
\begin{align}
&\rho_{12}=\frac{1}{\sigma_1\sigma_2}\int(x_1-\bar{x}_1)(x_2-\bar{x}_2)P(x_1,x_2|x_3,x_4){\rm d}x_1{\rm d}x_2,\nonumber\\
&\rho_{13}=\frac{1}{\sigma_1\sigma_3}\int(x_1-\bar{x}_1)(x_3-\bar{x}_3)P(x_1,x_3|x_2,x_4){\rm d}x_1{\rm d}x_3,\nonumber\\
&\rho_{14}=\frac{1}{\sigma_1\sigma_4}\int(x_1-\bar{x}_1)(x_4-\bar{x}_4)P(x_1,x_4|x_2,x_3){\rm d}x_1{\rm d}x_4
\end{align}
Substituting the formulation of $x_1$, Eq. (\ref{fourx1}), into the previous equation, we get
\begin{align}
&\rho_{12}=\frac{1}{\sigma_1\sigma_2}(a_{12}\sigma_2^2+a_{13}\sigma_2\sigma_3\rho_{23}+a_{14}\sigma_2\sigma_4\rho_{24}),\nonumber\\
&\rho_{13}=\frac{1}{\sigma_1\sigma_3}(a_{12}\sigma_2\sigma_3\rho_{23}+a_{13}\sigma_3^2+a_{14}\sigma_3\sigma_4\rho_{34}),\nonumber\\
&\rho_{14}=\frac{1}{\sigma_1\sigma_4}(a_{12}\sigma_2\sigma_4\rho_{24}+a_{13}\sigma_3\sigma_4\rho_{34}+a_{14}\sigma_4^2),
\end{align}
which drive
\begin{align}
\label{foura}
&a_{12}=\frac{\rho_{12}(1-\rho_{34}^2)-(\rho_{13}\rho_{23}+\rho_{14}\rho_{24})+(\rho_{13}\rho_{24}+\rho_{14}\rho_{23})\rho_{34}}
{1-\rho_{23}^2-\rho_{24}^2-\rho_{34}^2+2\rho_{23}\rho_{24}\rho_{34}}\frac{\sigma_1}{\sigma_2},\nonumber\\
&a_{13}=\frac{\rho_{13}(1-\rho_{24}^2)-(\rho_{12}\rho_{23}+\rho_{14}\rho_{34})+(\rho_{12}\rho_{34}+\rho_{14}\rho_{23})\rho_{24}}
{1-\rho_{23}^2-\rho_{24}^2-\rho_{34}^2+2\rho_{23}\rho_{24}\rho_{34}}\frac{\sigma_1}{\sigma_3},\nonumber\\
&a_{14}=\frac{\rho_{14}(1-\rho_{23}^2)-(\rho_{12}\rho_{24}+\rho_{13}\rho_{34})+(\rho_{12}\rho_{34}+\rho_{13}\rho_{24})\rho_{23}}
{1-\rho_{23}^2-\rho_{24}^2-\rho_{34}^2+2\rho_{23}\rho_{24}\rho_{34}}\frac{\sigma_1}{\sigma_4}.
\end{align}
Similarly, $c_1$ can be obtained by the definition of the variance of $x_1$:
\begin{equation}
\sigma_1^2=\int(x_1-\bar{x}_1)^2P(x_1|x_2, x_3, x_4){\rm d}x_1,
\end{equation}
which implies
\begin{equation}
\label{fourc1}
c_1^2=1-\frac{1}{\sigma_1^2}(a_{12}^2\sigma_2^2+a_{13}^2\sigma_3^2+a_{14}^2\sigma_4^2+2a_{12}a_{13}\sigma_2\sigma_3\rho_{23}+2a_{12}a_{14}\sigma_2\sigma_4\rho_{24}
+2a_{13}a_{14}\sigma_3\sigma_4\rho_{34}),
\end{equation}
where $V(r_1)=\sigma_1^2$ was used. Inserting Eq. (\ref{foura}) into Eq. (\ref{fourc1}) provides
\begin{multline}
\label{fourc}
c_1=\left[1-\rho_{23}^2-\rho_{24}^2-\rho_{34}^2+2\rho_{23}\rho_{24}\rho_{34}\right]^{-1/2}\left[1-\rho_{23}^2-\rho_{24}^2-\rho_{34}^2
+2\rho_{23}\rho_{24}\rho_{34}-\rho_{12}^2(1-\rho_{34}^2)\right.\\
-\rho_{13}^2(1-\rho_{24}^2)-\rho_{14}^2(1-\rho_{23}^2)+2\rho_{12}\rho_{13}(\rho_{23}-\rho_{24}\rho_{34})
+2\rho_{13}\rho_{14}(\rho_{34}-\rho_{23}\rho_{24})\\
\left.+2\rho_{12}\rho_{14}(\rho_{24}-\rho_{23}\rho_{34})\right]^{1/2}.
\end{multline}

Through the above analysis of simple cases, genera expressions for the coefficients that specify the independent and correlated sections separated from an arbitrary variable $x_i$ then can be derived, that is Eqs. (\ref{coefficientcorrelated}) and (\ref{coefficientuncorrelated}).

\section{Detailed calculation for test case 2}\label{appendixtestcase2}

\subsection{First-order contributions}
In the first nonlinear example, defined by Eq. (\ref{testcase2}), input variables are normally distributed with zero mean and covariance matrix indicated by Eq. (\ref{testcase2covariancematrix}). With Eq. (\ref{vcorrelated}), fractional variance contribution produced by each input factor alone are represented as
\begin{align}
V_1&=(\frac{\partial y}{\partial x_1})^2(\{\mu\})\cdot M_2(x_1)=4M_2(x_1),\label{firstorderx1} \\
V_2&=\frac{1}{2!\cdot 2!}(\frac{\partial^2 y}{\partial x_2^2})^2(\{\mu\})\cdot \left [M_4(x_2)-M_2^2(x_2)\right]=M_4(x_2)-M_2^2(x_2)\label{firstorderx2},\\
V_3&=(\frac{\partial y}{\partial x_3})^2(\{\mu\})\cdot M_2(x_3)=0,\label{firstorderx3}
\end{align}
To determine the independent, correlated and coupling effects that are contained in $V_1$, $x_1$ should be reformed on the bases of $x_2$ and $x_3$ as
\begin{equation}
x_1=a_{12}x_2+a_{13}x_3+c_1r_1,\label{reformx1}
\end{equation}
in which $r_1$ follows the same distribution with $x_1$, and coefficients $\{a_{12}, a_{13}, c_1\}$ are presented in  Eqs. (\ref{threea}) and (\ref{threec}). Substituting the formulation of $x_1$ into Eq. (\ref{firstorderx1}) yields
\begin{equation}
\begin{split}
V_1&=4E\left[(a_{12}x_2+a_{13}x_3+c_1r_1)^2\right]\\
&=4(a_{12}^2+a_{13}^2+2a_{12}a_{13}\rho_{23}+c_1^2)\\
&=4(1-c_1^2)+4c_1^2,
\end{split}
\end{equation}
which suggests a vanishing coupling effect but existent independent and correlated ones:
\begin{equation}
V_1^{\rm U}=4c_1^2, \qquad V_1^{\rm C}=4(1-c_1^2), \qquad V_1^{\rm UC}=0.
\end{equation}

Analogously, $x_2$ can be reformed as below based on $x_1$ and $x_3$ for the quantification of independent, correlated and coupling variance contributions contained in $V_2$:
\begin{equation}
x_2=a_{21}x_1+a_{23}x_3+c_2r_2,\label{reformx2}
\end{equation}
with $r_2$ holding the same distribution with $x_2$ and coefficients $\{a_{21}, a_{23}, c_2\}$ determined by Eqs. (\ref{coefficientcorrelated}) and (\ref{coefficientuncorrelated}). Inserting Eq. (\ref{reformx2}) into Eq. (\ref{firstorderx2}) drives
\begin{equation}
\begin{split}
V_2
=&E\left[(a_{21}x_1+a_{23}x_3+c_2r_2)^4\right]-E^2\left[(a_{21}x_1+a_{23}x_3+c_2r_2)^2\right]\\
=&E\left[(a_{21}^2x_1^2+a_{23}^2x_3^2+2a_{21}a_{23}x_1x_3+2a_{21}c_2x_1r_2+2a_{23}c_2x_3r_2+c_2^2r_2^2)^2\right]\\
&-E^2\left[(a_{21}x_1+a_{23}x_3+c_2r_2)^2\right]\\
=&\left[3(a_{21}^4+a_{23}^4)+6a_{21}^2a_{23}^2(2\rho_{13}^2+1)+12a_{21}a_{23}\rho_{13}(a_{21}^2+a_{23}^2)\right]\\
&+\left[6c_2^2(a_{21}^2+a_{23}^2+2a_{21}a_{23}\rho_{13})\right]+3c_2^4-\left[(a_{21}^2+a_{23}^2+2a_{21}a_{23}\rho_{13})+c_2^2\right]^2\\
=&2(1-c_2^2)^2+2c_2^4+4c_2^2(1-c_2^2),
\end{split}
\end{equation}
where the first term is produced by the correlations of $x_2$ with the remaining inputs, the second one by its independence, and the third one by the coupling effect associated with correlations and independence, indicating
\begin{equation}
V_2^{\rm U}=2c_2^4,\qquad
V_2^{\rm C}=2(1-c_2^2)^2,\qquad
V_2^{\rm UC}=4c_2^2(1-c_2^2).
\end{equation}

\subsection{Second-order contributions}
The second-order partial variance contributions produced by the combinations between each pair of inputs are expressed as
\begin{align}
V_{12}&=\frac{1}{2!\cdot 2!}(\frac{\partial^3 y}{\partial x_1^2 \partial x_2})^2(\{\mu\}) \cdot {\rm cov}(x_1^4, x_2^2)+\frac{2}{2!}(\frac{\partial y}{\partial x_1}\cdot \frac{\partial^3 y}{\partial x_1^2 \partial x_2})(\{\mu\})\cdot {\rm cov}(x_1^3, x_2)\nonumber\\
&=16\left[{\rm cov}(x_1^4, x_2^2)+{\rm cov}(x_1^3, x_2)\right],\label{secondorder12}\\
V_{13}&=(\frac{\partial^2 y}{\partial x_1\partial x_3})^2(\{\mu\})\cdot \left[{\rm cov}(x_1^2, x_3^2)-{\rm cov}^2(x_1, x_3)\right]\nonumber\\
&={\rm cov}(x_1^2, x_3^2)-\rho_{13}^2,\label{secondorder13} \\
V_{23}&=0.\label{secondorder23}
\end{align}
Considering the reforming of $x_1$ on the bases of $x_2$ and $x_3$, the partial variance contribution $V_{12}$ is determined as
\begin{equation}
\begin{split}
V_{12}
=&16E\left [x_2^2 (a_{12}x_2+a_{13}x_3+c_1r_1)^4\right ]+16E\left [x_2 (a_{12}x_2+a_{13}x_3+c_1r_1)^3\right ]\\
=&16E\left [x_2^2(a_{12}^2x_2^2+a_{13}^2x_3^2+2a_{12}a_{13}x_2x_3+2a_{12}c_1x_2r_1+2a_{13}c_1x_3r_1+c_1^2r_1^2)^2\right] \\
&+16E\left [x_2(a_{12}^3x_2^3+a_{13}^3x_3^3+3a_{12}^2a_{13}x_2^2x_3+3a_{12}a_{13}^2x_2x_3^2)\right]\\
&+16E\left [x_2(3c_1(a_{12}^2x_2^2+a_{13}^2x_3^2+2a_{12}a_{13}x_2x_3)r_1+3c_1^2(a_{12}x_2+a_{13}x_3)r_1^2+c_1^3r_1^3)\right]\\
=&48(1-c_1^2)(1-c_1^2+\rho_{12}+4\rho_{12}^2)+48c_1^4+48c_1^2(2-2c_1^2+\rho_{12}+4\rho_{12}^2),
\end{split}
\end{equation}
which, from the first fraction to the third one, are separately produced by the correlations of $x_1$ with the rest, the independence of $x_1$, and the coupling effect between correlations and independence, specifying
\begin{align}
&V_{12}^{{\rm C}_1}=48(1-c_1^2)(1-c_1^2+\rho_{12}+4\rho_{12}^2),\nonumber \\
&V_{12}^{{\rm U}_1}=48c_1^4,\qquad
V_{12}^{{\rm UC}_1}=48c_1^2(2-2c_1^2+\rho_{12}+4\rho_{12}^2).
\end{align}

Regarding the formulation of input $x_2$, $V_{12}$ can be also calculated as
\begin{equation}
\begin{split}
V_{12}&=16\left [x_1^4(a_{21}x_1+a_{23}x_3+c_2r_2)^2\right ]+16E\left [x_1^3 (a_{21}x_1+a_{23}x_3+c_2r_2)\right ]\\
&=16\left [15a_{21}^2+a_{23}^2(12\rho_{13}^2+3)+30a_{21}a_{23}\rho_{13}+3a_{21}+3a_{23}\rho_{13}\right ]+48c_2^2 \\
&=48(1-c_2^2+\rho_{12}+4\rho_{12}^2)+48c_2^2,
\end{split}
\end{equation}
which suggests a vanishing coupling effect but existent correlated (first term with brackets) and independent (last term) ones:
\begin{eqnarray}
V_{12}^{{\rm U}_2}=48c_2^2,\qquad
V_{12}^{{\rm C}_2}=48(1-c_2^2+\rho_{12}+4\rho_{12}^2),\qquad
V_{12}^{{\rm UC}_2}=0.
\end{eqnarray}

Recalling the reforming of $x_1$ on the bases of $x_2$ and $x_3$, the partial variance contribution $V_{13}$ can be analogously obtained as
\begin{equation}
\begin{split}
V_{13}&=E\left [x_3^2\cdot (a_{12}x_2+a_{13}x_3+c_1r_1)^2\right ]-\rho_{13}^2 \\
&=a_{12}^2\left(2\rho_{23}^2+1\right)+3a_{13}^2+6a_{12}a_{13}\rho_{23}-\rho_{13}^2+c_1^2\\
&=1-c_1^2+2a_{12}\rho_{23}(a_{13}+a_{12}\rho_{23})+2a_{13}(a_{13}+a_{12}\rho_{23})-\rho_{13}^2+c_1^2\\
&=1-c_1^2+\rho_{13}^2+c_1^2,
\end{split}
\end{equation}
which only contains the correlated effect (first three components) contributed by the correlations of $x_1$ with the rest variables, and independent one (last component) produced by the independence of $x_1$:
\begin{equation}
V_{13}^{{\rm U}_1}=c_1^2, \qquad V_{13}^{{\rm C}_1}=1-c_1^2+\rho_{13}^2, \qquad V_{13}^{{\rm UC}_1}=0.
\end{equation}
$V_{13}$ can be equivalently obtained by reforming $x_3$ on the bases of $x_1$ and $x_2$, constituted of
\begin{equation}
V_{13}^{{\rm U}_3}=c_3^2, \qquad V_{13}^{{\rm C}_3}=1-c_3^2+\rho_{13}^2, \qquad V_{13}^{{\rm UC}_3}=0.
\end{equation}

\subsection{Third-order contributions}
The third-order partial variance contribution associated with the second test case is expressed as
\begin{equation}
\label{thirdorder123}
\begin{split}
V_{123}&=\frac{2}{2!}(\frac{\partial^2 y}{\partial x_2^2}\cdot \frac{\partial ^2 y}{\partial x_1\partial x_3})(\{\mu\})\left[{\rm cov}(x_1, x_2^2, x_3)-M_2(x_2){\rm cov}(x_1, x_3)\right]\\
&={\rm cov}(x_1,x_2^2, x_3)-M_2(x_2)\rho_{13}.
\end{split}
\end{equation}
By introducing the formulation of $x_2$ on the bases of $x_1$ and $x_3$ (Eq. (\ref{reformx2})), we get
\begin{equation}
\begin{split}
V_{123}
=&E\left[x_1x_3(a_{21}x_1+a_{23}x_3+c_2r_2)^2\right ]-E\left[(a_{21}x_1+a_{23}x_3+c_2r_2)^2\right]\rho_{13}\\
=&E\left[x_1x_3(a_{21}x_1+a_{23}x_3)^2\right ]-E\left[(a_{21}x_1+a_{23}x_3)^2\right]\rho_{13}+E\left[c_2^2x_1x_3r_2^2\right]-E\left[c_2^2r_2^2\right]\rho_{13},
\end{split}
\end{equation}
where the first two items are contributed by the correlations of $x_2$ with both $x_1$ and $x_3$, and the last two items, summing to zero, are provided by the coupling effect between the independence of $x_2$ and the correlation of $x_1$ with $x_3$. $V_{123}$ is then computed as
\begin{equation}
\begin{split}
V_{123}&=2a_{21}^2\rho_{13}+2a_{23}^2\rho_{13}+2a_{21}a_{23}\rho_{13}^2+2a_{21}a_{23}\\
&=2a_{21}\rho_{13}(a_{21}+a_{23}\rho_{13})+2a_{23}(a_{21}+a_{23}\rho_{13})\\
&=2a_{21}\rho_{12}\rho_{13}+2a_{23}\rho_{12}\\
&=2\rho_{12}\rho_{23},
\end{split}
\end{equation}
which is totally contributed by input correlations:
\begin{equation}
V_{123}^{{\rm U}_2}=V_{123}^{{\rm UC}_2}=0, \qquad V_{123}^{{\rm C}_2}=2\rho_{12}\rho_{23}.
\end{equation}
We can analogously state that
\begin{equation}
V_{123}^{{\rm U}_1}=V_{123}^{{\rm UC}_1}=V_{123}^{{\rm U}_3}=V_{123}^{{\rm UC}_3}=0, \qquad V_{123}^{{\rm C}_1}=V_{123}^{{\rm C}_3}=2\rho_{12}\rho_{23}.
\end{equation}

\section{The derivation of cov($x_1, x_2, x_3, x_4$)}\label{appendixtestcase3}

In the second nonlinear test case, the covariance among four correlated variables is involved:
\begin{equation}
{\rm cov}(x_1, x_2, x_3, x_4)=E\left[(x_1-\mu_1)(x_2-\mu_2)(x_3-\mu_3)(x_4-\mu_4)\right],\label{fourthcovariance}
\end{equation}
where $(x_1, x_2, x_3, x_4)\sim N(\bm{\mu}, \Sigma)$ with mean vector $\bm{\mu}=(1,2,2,1)$ and covariance matrix displayed by Eq. (\ref{testcase3covariance}). Select $x_1$ to be formulated on the bases of the rest:
\begin{equation}
x_1=a_{12}x_2+a_{13}x_3+a_{14}x_4+c_1r_1,\label{fourthx1}
\end{equation}
with coefficients $\{a_{12}, a_{13}, a_{14}, c_1\}$ presented in Eqs. (\ref{foura}) and (\ref{fourc}), and $r_1$ holding the same distribution as $x_1$. Inserting Eq. (\ref{fourthx1}) into Eq. (\ref{fourthcovariance}) provides
\begin{equation}
\begin{split}
{\rm cov}(x_1, x_2, x_3, x_4)
&=E\left[(a_{12}x_2+a_{13}x_3+a_{14}x_4+c_1r_1)x_2x_3x_4\right]\\
&=E\left[a_{12}x_2^2x_3x_4\right]+E\left[a_{13}x_2x_3^2x_4\right]+E\left[a_{14}x_2x_3x_4^2\right].\label{deffourthcovaraince}
\end{split}
\end{equation}
For the evaluation of $E\left[a_{12}x_2^2x_3x_4\right]$, another variable should also be formulated based on the other variables enclosed in the brackets. Here we formulate $x_3$ (equivalent to formulate $x_2$ or $x_4$ as input variables are normally distributed) as
\begin{equation}
x_3=a_{32}x_2+a_{34}x_4+c_3r_3,\label{reformx3}
\end{equation}
where $r_3$ holds the same distribution as $x_3$ and coefficients $\{a_{32}, a_{34}, c_3\}$ are determined through Eqs. (\ref{coefficientcorrelated}) and (\ref{coefficientuncorrelated}). Then we can obtain
\begin{equation}
\begin{split}
E\left[a_{12}x_2^2x_3x_4\right]&=\left[a_{12}a_{32}x_2^3x_4+a_{12}a_{34}x_2^2x_4^2\right]\\
&=2a_{12}\rho_{24}(a_{32}+a_{34}\rho_{24})+a_{12}(a_{34}+a_{32}\rho_{24})\\
&=2a_{12}\rho_{23}\rho_{24}+a_{12}\rho_{34}.
\end{split}
\end{equation}
The rest two average items on the right side of Eq. (\ref{deffourthcovaraince}) can be similarly determined as
\begin{align}
E\left[a_{13}x_2x_3^2x_4\right]&=2a_{13}\rho_{23}\rho_{34}+a_{13}\rho_{24},\\
E\left[a_{14}x_2x_3x_4^2\right]&=2a_{14}\rho_{24}\rho_{34}+a_{14}\rho_{23}.
\end{align}
We now have
\begin{equation}
\begin{split}
{\rm cov}(x_1, x_2, x_3,x_4)
=&2(a_{12}\rho_{23}\rho_{24}+a_{13}\rho_{23}\rho_{34}+a_{14}\rho_{24}\rho_{34})+(a_{12}\rho_{34}+a_{13}\rho_{24}+a_{14}\rho_{23})\\
=&\rho_{34}(a_{12}+a_{13}\rho_{23}+a_{14}\rho_{24})+\rho_{24}(a_{13}+a_{12}\rho_{23}+a_{14}\rho_{34})\\
&+\rho_{23}(a_{14}+a_{12}\rho_{24}+a_{13}\rho_{34})\\
=&\rho_{12}\rho_{34}+\rho_{13}\rho_{24}+\rho_{14}\rho_{23}.
\end{split}
\end{equation}

\section{Ishigami function}\label{ishigamifunction2}
The Ishigami function is defined by Eq. (\ref{ishigamifunction}) with input variables uniformly distributed in the interval $[-\pi, \pi]$ which provides $\mu_i$=0 and $\sigma_i=\pi/\sqrt{3}$ for $i\in\{1, 2, 3\}$.

\subsection{First-order contributions}
With help of Eq. (\ref{vcorrelated}), the main partial variance contribution produced by input $x_1$ alone is represented as
\begin{equation}
\label{ishigamainx1}
V_1=\sum_{i,j=0}^{\infty}\frac{1}{i!\cdot j!}(\frac{\partial^i y}{\partial x_1^i}\cdot \frac{\partial^jy}{\partial x_1^j})(\{\mu\})\cdot \left[M_{i+j}(x_i)-M_i(x_1)M_j(x_1)\right].
\end{equation}
The existent partial derivatives of $y$ with respect to $x_1$ are totally provided by the first term of Eq. (\ref{ishigamifunction}): $\sin(x_1)$ as $\mu_3=0$. Zero mean value of $x_1$ determines
\begin{equation}
\label{derivativesinx1}
\left(\left.\frac{\partial^i \sin(x_1)}{\partial x_1^i}\right|_{\mu_1=0}\right)=\left\{
\begin{array}{ll}
(-1)^{(i-1)/2}, & \text{$i$ is odd},\\
0, & \text{$i$ is even}.
\end{array}
\right.
\end{equation}
Substituting Eqs. (\ref{centralmomentuniform}) and (\ref{derivativesinx1}) into Eq. (\ref{ishigamainx1}) yields
\begin{equation}
V_1=\sum_{i,j=0}^{\infty}\frac{(-1)^{i+j}}{(2i+1)!(2j+1)!}\cdot\frac{\pi^{2(i+j+1)}}{2i+2j+3}=0.5.
\end{equation}

Analogously, the main variance contribution produced by $x_2$ is represented as
\begin{equation}
\label{ishigamimainx2}
V_2=\sum_{i,j=0}^{\infty}\frac{1}{i!\cdot j!}(\frac{\partial^i y}{\partial x_2^i}\cdot \frac{\partial^jy}{\partial x_2^j})(\{\mu\})\cdot\left[M_{i+j}(x_2)-M_i(x_2)M_j(x_2)\right],
\end{equation}
which is provided by the second term of Eq. (\ref{ishigamifunction}): $7\sin^2(x_2)$. Zero mean value of $x_2$ suggests
\begin{equation}
\label{derivativesinx2}
\left(\left.\frac{\partial^i \sin^2(x_2)}{\partial x_2^i}\right|_{\mu_2=0}\right)=\left\{
\begin{array}{ll}
2^{i-1}(-1)^{(i+2)/2}, & \text{$i$ is even},\\
0, & \text{$i$ is odd}.
\end{array}
\right.
\end{equation}
By inserting Eqs. (\ref{centralmomentuniform}) and (\ref{derivativesinx2}) into Eq. (\ref{ishigamimainx2}), we get
\begin{equation}
V_2=7^2\sum_{i,j=0}^{\infty}\frac{(-1)^{i+j}4^{i+j+1}\pi^{2(i+j+2)}}{(2i+2)!(2j+2)!}\left[\frac{1}{2i+2j+5}-\frac{1}{(2i+3)(2j+3)}\right]=6.125.
\end{equation}
Zero mean value of $x_1$ also suggests a vanishing main partial variance contribution produced by $x_3$ alone since $x_3$ just appears in the combination with $\sin$ function of $x_1$ in the form of Ishigami function. $V_1$ and $V_2$ are, respectively, embodied by $\sin$ functions of $x_1$ and $x_2$ which explain the vanishing independent and correlated effects but existent coupling one, that is
\begin{equation}
V_1^{\rm U}=V_1^{\rm C}=V_2^{\rm U}=V_2^{\rm C}=0, \qquad V_1^{\rm UC}=0.5, \qquad V_2^{\rm UC}=6.125.
\end{equation}

\subsection{Second-order contributions}
For partial variance contributions of second-order, the form of Ishigami function provides $V_{12}=V_{23}=0$ and
\begin{equation}
\label{appendixishigamiv13}
\begin{split}
V_{13}
=&0.1^2\sum_{i,j=0}^{\infty}\frac{1}{i!\cdot j!\cdot4!\cdot4!}(\frac{\partial^{4+i}y}{\partial x_1^i\partial x_3^4}\cdot \frac{\partial^{4+j}y}{\partial x_1^j\partial x_3^4})(\{\mu\})\cdot\left[{\rm cov}(x_1^{i+j}, x_4^8)-{\rm cov}(x_1^i, x_3^4){\rm cov}(x_1^j, x_3^4)\right]\\
&+2\cdot 0.1\sum_{i, j=0}^{\infty}\frac{1}{i!\cdot j!\cdot4!}(\frac{\partial^iy}{\partial x_1^i}\cdot \frac{\partial^{4+j}y}{\partial x_1^j\partial x_3^4})(\{\mu\})\cdot\left[{\rm cov}(x_1^{i+j}, x_3^4)-M_i(x_1){\rm cov}(x_1^j, x_3^4)\right]\\
=&0.1^2\sum_{i,j=0}^{\infty}\frac{1}{i!\cdot j!}(\frac{\partial^i\sin(x_1)}{\partial x_1^i}\cdot \frac{\partial^j\sin(x_1)}{\partial x_1^j})(\mu_1=0)\cdot\left[{\rm cov}(x_1^{i+j}, x_4^8)-{\rm cov}(x_1^i, x_3^4){\rm cov}(x_1^j, x_3^4)\right]\\
&+2\cdot 0.1\sum_{i, j=0}^{\infty}\frac{1}{i!\cdot j!}(\frac{\partial^i\sin(x_1)}{\partial x_1^i}\cdot \frac{\partial^{j}\sin(x_1)}{\partial x_1^j})(\mu_1=0)\cdot\left[{\rm cov}(x_1^{i+j}, x_3^4)-M_i(x_1){\rm cov}(x_1^j, x_3^4)\right]
\end{split}
\end{equation}
By substituting Eq. (\ref{derivativesinx1}), $V_{13}$ is simplified as
\begin{equation}
V_{13}=0.1\sum_{i,j=0}^{\infty}\frac{(-1)^{i+j}}{(2i+1)!(2j+1)!}\left[0.1{\rm cov}(x_1^{2i+2j+2}, x_3^8)+2{\rm cov}(x_1^{2i+2j+2}, x_3^4)\right].
\end{equation}
To determine $V_{13}$, it is necessary to derive the covariance items enclosed in the above square parentheses.

Firstly, consider the generation of $x_1$ on the basis of $x_3$:
\begin{equation}
\begin{split}
x_1&=\rho_{13}\frac{\sigma_1}{\sigma_3}x_3+\sqrt{1-\rho_{13}^2}r_1\\
&=\rho_{13}x_3+\sqrt{1-\rho_{13}^2}r_1,
\end{split}
\end{equation}
with $r_1$ holding the same distribution as $x_1$. Inserting the above expression into covariance items contained in Eq. (\ref{appendixishigamiv13}) yields
\begin{equation}
\label{ishigamireformx11}
\begin{split}
{\rm cov}&(x_1^{2(i+j+1)}, x_3^8)\\
&=E\left[x_3^{8}\sum_{k=0}^{i+j+1}\binom{2(i+j+1)}{2k}\rho_{13}^{2k}(1-\rho_{13}^2)^{i+j+1-k}x_3^{2k}r_1^{2i+2j+2-2k}\right]\\
&=\sum_{k=0}^{i+j+1}\binom{2(i+j+1)}{2k}\rho_{13}^{2k}(1-\rho_{13}^2)^{i+j+1-k}M_{2i+2j+2-2k}(x_1)M_{8+2k}(x_3)\\
&=\sum_{k=0}^{i+j+1}\binom{2(i+j+1)}{2k}\cdot \frac{\pi^{2(i+j+1)+8}\rho_{13}^{2k}(1-\rho_{13}^2)^{i+j+1-k}}{(2(i+j+1)-2k+1)(8+2k+1)},
\end{split}
\end{equation}
and
\begin{equation}
\label{ishigamireformx12}
\begin{split}
{\rm cov}&(x_1^{2(i+j+1)}, x_3^4)\\
&=E\left[x_3^{4}\sum_{k=0}^{i+j+1}\binom{2(i+j+1)}{2k}\rho_{13}^{2k}(1-\rho_{13}^2)^{i+j+1-k}x_3^{2k}r_1^{2i+2j+2-2k}\right]\\
&=\sum_{k=0}^{i+j+1}\binom{2(i+j+1)}{2k}\rho_{13}^{2k}(1-\rho_{13}^2)^{i+j+1-k}M_{2(i+j+1)-2k}(x_1)M_{4+2k}(x_3)\\
&=\sum_{k=0}^{i+j+1}\binom{2(i+j+1)}{2k}\cdot \frac{\pi^{2(i+j+1)+4}\rho_{13}^{2k}(1-\rho_{13}^2)^{i+j+1-k}}{(2(i+j+1)-2k+1)(4+2k+1)}.
\end{split}
\end{equation}
$V_{13}$ is now obtained as
\begin{multline}
V_{13}=0.1\sum_{i,j=0}^{\infty}\sum_{k=0}^{i+j+1}\frac{(-1)^{i+j}\pi^{2(i+j+1)+4}}{(2i+1)!(2j+1)!}\cdot \binom{2(i+j+1)}{2k}\\ \times\frac{\rho_{13}^{2k}(1-\rho_{13}^2)^{i+j+1-k}}{2(i+j-k)+3}\left[\frac{0.1\cdot \pi^{4}}{9+2k}+\frac{2}{5+2k}\right].
\end{multline}
The independent and correlated partial variance contributions included in $V_{13}$ can be specified by setting $k=0$ and $k=2(i+j+1)$, respectively. We get
\begin{align}
V_{13}^{\rm U_1}
&=0.1\sum_{i,j=0}^{\infty}\frac{(-1)^{i+j}\pi^{2(i+j+1)+4}}{(2i+1)!(2j+1)!}\frac{(1-\rho_{13}^2)^{i+j+1}}{2(i+j)+3}\left[\frac{0.1\cdot \pi^{4}}{9}+\frac{2}{5}\right],\\
V_{13}^{\rm C_1}
&=0.1\sum_{i,j=0}^{\infty}\frac{(-1)^{i+j}\pi^{2(i+j+1)+4}}{(2i+1)!(2j+1)!}\rho_{13}^{2(i+j+1)}\left[\frac{0.1\cdot \pi^{4}}{2(i+j)+11}+\frac{2}{2(i+j)+7}\right].
\end{align}
The partial variance contribution provided by the coupling effect between the independent and correlated sections is then spontaneously determined for $x_1$ by
\begin{equation}
V_{13}^{\rm UC_1}=V_{13}-V_{13}^{\rm U_1}-V_{13}^{\rm C_1}.
\end{equation}

Optionally, $x_3$ can be formulated in terms of $x_1$ as:
\begin{equation}
\begin{split}
x_3&=\rho_{13}\frac{\sigma_3}{\sigma_1}x_1+\sqrt{1-\rho_{13}^2}r_3\\
&=\rho_{13}x_1+\sqrt{1-\rho_{13}^2}r_3,
\end{split}
\end{equation}
with $r_3$ holding the same distribution as $x_3$. Then we get
\begin{equation}
\begin{split}
{\rm cov}&(x_1^{2(i+j+1)}, x_3^8)\\
&=E\left[x_1^{2(i+j+1)}\sum_{k=0}^4\binom{8}{2k}\rho_{13}^{2k}(1-\rho_{13}^2)^{4-k}x_1^{2k}r_3^{8-2k}\right]\\
&=\sum_{k=0}^4\binom{8}{2k}\rho_{13}^{2k}(1-\rho_{13}^2)^{4-k}M_{2(i+j+1)+2k}(x_1)M_{8-2k}(x_3)\\
&=\sum_{k=0}^4\binom{8}{2k}\cdot \frac{\pi^{2(i+j+1)+8}\rho_{13}^{2k}(1-\rho_{13}^2)^{4-k}}{(2(i+j+1)+2k+1)(8-2k+1)},
\end{split}
\end{equation}
and
\begin{equation}
\begin{split}
{\rm cov}&(x_1^{2(i+j+1)}, x_3^4)\\
&=E\left[x_1^{2(i+j+1)}\sum_{l=0}^2\binom{4}{2l}\rho_{13}^{2l}(1-\rho_{13}^2)^{2-l}x_1^{2l}r_3^{4-2l}\right]\\
&=\sum_{l=0}^2\binom{4}{2l}\rho_{13}^{2l}(1-\rho_{13}^2)^{2-l}M_{2(i+j+1)+2l}(x_1)M_{4-2l}(x_3)\\
&=\sum_{l=0}^2\binom{4}{2l}\cdot \frac{\pi^{2(i+j+1)+4}\rho_{13}^{2l}(1-\rho_{13}^2)^{2-l}}{(2(i+j+1)+2l+1)(4-2l+1)},
\end{split}
\end{equation}
which yield
\begin{multline}
V_{13}=0.1\sum_{i,j=0}^{\infty}\frac{(-1)^{i+j}\pi^{2(i+j+1)+4}}{(2i+1)!(2j+1)!}\left[0.1\sum_{k=0}^4\binom{8}{2k}\cdot \frac{\pi^{4}\rho_{13}^{2k}(1-\rho_{13}^2)^{4-k}}{(2i+2j+2k+3)(9-2k)}\right.\\
\left.+2\sum_{l=0}^2\binom{4}{2l}\cdot \frac{\rho_{13}^{2l}(1-\rho_{13}^2)^{2-l}}{(2i+2j+2l+3)(5-2l)}\right].
\end{multline}
The Independent and correlated effects contained in the above partial variance contribution can be, separately, calculated by setting $k=0,l=0$ and $k=4,l=2$, specifying
\begin{align}
V_{13}^{\rm U_3}
&=0.1\sum_{i,j=0}^{\infty}\frac{(-1)^{i+j}\pi^{2(i+j+1)+4}}{(2i+1)!(2j+1)!}\left[\frac{0.1\cdot \pi^{4}(1-\rho_{13}^2)^4}{9(2i+2j+3)}+\frac{2 (1-\rho_{13}^2)^2}{5(2i+2j+3)}\right],\\
V_{13}^{\rm C_3}
&=0.1\sum_{i,j=0}^{\infty}\frac{(-1)^{i+j}\pi^{2(i+j+1)+4}}{(2i+1)!(2j+1)!}\left[\frac{0.1\cdot \pi^{4}\rho_{13}^8}{2i+2j+11}+\frac{2 \rho_{13}^4}{2i+2j+7}\right].
\end{align}
The coupling variance contribution is then naturally determined for $x_3$ by
\begin{equation}
V_{13}^{\rm UC_3}=V_{13}-V_{13}^{\rm U_3}-V_{13}^{\rm C_3}.
\end{equation}


\begin{thebibliography}{99}

\bibitem{Mroz1987} T.A. Mroz Econometrica, {\bf 55} (1987) 765--799.

\bibitem{Ligmann2014} A. Ligmann-Zielinska, D.B. Kramer, K.S. Cheruvelil, P.A. Soranno, PloS one {\bf 9} (2014) e109779.

\bibitem{Becker2011} W. Becker, J. Rowson, J.E. Oakley, A. Yoxall, G. Manson, K. Worden, J. Biomech. {\bf 44} (2011) 1499--506.

\bibitem{Pannell1997} D.J. Pannell, Agric. Econ. {\bf 16} (1997) 139--152.

\bibitem{Saltelli2005} A. Saltelli, M. Ratto, S. Tarantola, F. Campolongo, Chem. rev. {\bf 105} (2005) 2811--28.

\bibitem{Carr1993} S. Carr, G.J. Savage, in: T. Lee (Ed.), Mathematical Computation with Maple V: Ideas and Applications, Boston, 1993, pp. 118--127.

\bibitem{Morio2011} J. Morio, Eur. J. Phys. {\bf 32} (2011) 1577-83.

\bibitem{Timme2014} M. Timme, J. Casadiego, J. Phys. A: Math. Theor. {\bf 47} (2014) 343001.

\bibitem{Posselt2016} D.J. Posselt, B. Fryxell, A. Molod, B. Williams, J. Climate {\bf 29} (2016) 455--479.

\bibitem{Guo2013} L. Guo, Z. Luo, Y. Zhu, J. Stat. Mech. {\bf 2013} (2013) 11013.

\bibitem{Zhu2013} Y. Zhu, W. Li, X. Cai, Physica A {\bf 392} (2013) 6596--602.

\bibitem{Cacuci2005} D.G. Cacuci, M. Ionescu-Bujor, I.M. Navon, Sensitivity and Uncertainty Analysis: Applications to Large-Scale Systems, Vol. 2, CRC press, London, 2005.

\bibitem{Frankel2010} J.I. Frankel, M. Keyhani, M. G. Huang, Appl. Math. Comput. {\bf 217} (2010) 363-375.

\bibitem{Farina2013} D. Farina, P. Hammes, S. Reitzinger, Appl. Math. Comput. {\bf 219} (2013) 7181-7186.

\bibitem{Pedro2016} S.A. Pedro, H.E.Z. Tonnang, S. Abelman, Appl. Math. Comput. {\bf 279} (2016) 170-186.

\bibitem{Iooss2015} B. Iooss, P. Lema\^itre, in: G. Dellino, C. Meloni (Eds.), Uncertainty Management in Simulation-Optimization of Complex Systems, Operations Research/Computer Science Interfaces Series, {\bf 59}, Springer, Boston, 2015, P. 101--122.

\bibitem{zhu2017} Y. Zhu, Q.A. Wang, W. Li, X. Cai, Physica A {\bf 469} (2017) 52--59.

\bibitem{Sobol1993} I.M. Sobol', Math. Model. Comput. Exper. {\bf 1} (1993) 407--414.

\bibitem{Helton2006} J.C. Helton, J.D. Johnson, C.J. Sallaberry, C.B. Storlie, Reliab. Eng. Syst. Safe. {\bf 91} (2006) 1175--209.

\bibitem{Kucherenko2015} S. Kucherenko, D. Albrecht, A. Saltelli, Exploring multi-dimensional spaces: a Comparison of Latin Hypercube and Quasi Monte Carlo Sampling Techniques arXiv:1505.02350, 2015.

\bibitem{Saltelli2001} A. Saltelli, M. Marco, S. Tarantola, in: P. Prado, R. Bolado (Eds.), 3rd Int. Symp. on Sensitivity Analysis of Model Output, Madrid, 2001, p~21--25.

\bibitem{Xu2008} C. Xu, G.Z. Gertner, Reliab. Eng. Syst. Safe. {\bf 93} (2008) 1563--73.

\bibitem{Most2012} T. Most, in: M. Vorechovsk\'y, V. Sad\'ilek, S. Seitl, V. Vesel\'y, R.L. Muhanna, R.L. Mullen (Eds.), Proc. 5th Int. Conf. on Reliable Engineering Computing, Brno, 2012, p~335--351.

\bibitem{Hao2013} W. Hao, Z. Lu, L. Li, Comput. Phys. Commun. {\bf 184} (2013) 1401--13.

\bibitem{Li2017} L. Li, Z. Lu, K. Zhang, Q. Gao, Aerosp. Sci. Technol. {\bf 62} (2017) 75--86.

\bibitem{Ishigami1990} T. Ishigami, T. Homma, in: Proc. 1st Int. Symp. on Uncertainty Modelling and Analysis, IEEE, 1990, p~398--403.

\bibitem{Lira2016} I. Lira, Meas. Sci. Technol. {\bf 27} (2016) 075006--13.

\bibitem{Fraser2009} C. Fraser, et al., science {\bf 324} (2009) 1557--61.

\bibitem{Holme2015} P. Holme, N. Masuda, PloS one {\bf 10} (2015) e0120567.

\bibitem{anderson1992} R.M. Anderson, R.M. May, B. Anderson, Infectious diseases of humans: dynamics and control, Vol. 28, Oxford university press, Oxford, 1992.

\bibitem{diekmann1990} O. Diekmann, J.A.P. Heesterbeek, J.A.J. Metz, J. Math. Biol., {\bf 28} (1990) 365-382.

\bibitem{Diekmann2010} O. Diekmann, J.A.P. Heesterbeek, M.G. Roberts, J. R. Soc. Interface {\bf 7} (2010) 873--885.

\bibitem{Macdonald1957} G. Macdonald, The Epidemiology and Control of Malaria, Oxford University Press, Oxford, 1957.

\bibitem{Van2002} P. Van den Driessche, J. Watmough, Math. Biosci. {\bf 180} (2002) 29--48.

\bibitem{Usman2016} I.G. Usman, M.A. Liman, I. Yusuf, S. Abdulrahman, Z. U. Garba, N. Isah, GARJPAS {\bf 5} (2016) 011--017.

\bibitem{Mukandavire2006} Z. Mukandavire, W. Garira, J. Biol. Syst. {\bf 14} (2006) 323--355.

\bibitem{Safiel2012} R. Safiel, E.S. Massawe, D.O. Makinde, Amer. J. Math. Stat. {\bf 2} (2012) 75--88.


\end{thebibliography}
\end{document}